\documentclass[preprint,showpacs,amsmath,amssymb,pra,superscriptaddress]{revtex4}

\usepackage{graphicx}
\usepackage{dcolumn}
\usepackage{bm}

\begin{document}


\title{Rydberg Atoms in a Magnetic Guide}
\date{\today}

\author{Igor Lesanovsky}
\email[]{ilesanov@physi.uni-heidelberg.de}
\affiliation{%
Physikalisches Institut, Universit\"at Heidelberg, Philosophenweg 12, 69120 Heidelberg, Germany}%
\author{J\"org Schmiedmayer}
\email[]{joerg.schmiedmayer@physi.uni-heidelberg.de}
\affiliation{%
Physikalisches Institut, Universit\"at Heidelberg, Philosophenweg 12, 69120 Heidelberg, Germany}%
\author{Peter Schmelcher$^{~\S,}$}
\email[]{Peter.Schmelcher@pci.uni-heidelberg.de}
\affiliation{%
Physikalisches Institut, Universit\"at Heidelberg, Philosophenweg 12, 69120 Heidelberg, Germany}%
\affiliation{%
Theoretische Chemie, Institut f\"ur Physikalische Chemie,
Universit\"at Heidelberg,
INF 229, 69120 Heidelberg, Germany}%
\thanks{Corresponding author}

\date{\today}

\begin{abstract}\label{txt:abstract}
We investigate electronically excited atoms in a magnetic guide. It turns out that the
Hamiltonian describing this system possesses a wealth of both unitary as well as antiunitary symmetries
that constitute an uncommon extensive symmetry group. One consequence is the two-fold
degeneracy of any energy level. The spectral properties are investigated for a wide
range of field gradients and the spatial distributions of the spin polarization
are analyzed. Wave lengths, oscillator strengths and selection rules are provided
for the corresponding electromagnetic transitions. The effects due to an additional
homogeneous bias field constituting a Ioffe-Pritchard trap are explored equally.
\end{abstract}

\maketitle
%
\section{Introduction}\label{sec:introduction}
%

External fields are nowadays widely used to control the motion
of atoms including their cooling and trapping as well as the preparation
of their internal states. Optical lattices and atom chips are two
major examples of devices that allow to deal with atomic ensembles
but also possess the perspective of manipulating single atoms for the purpose of
quantum information processing. To this end it is indispensable
to understand the structure and behaviour of (excited) individual atoms in traps.
In the case of the atom chip (see ref.\cite{Folman02} and references therein)
tight magnetic traps on the micrometer scale can be created exhibiting large
field gradients which are not accessible in the case of macrosopic traps.
Highly excited Rydberg atoms therefore start to 'feel' the variation i.e. the inhomogeneity of the
magnetic field across the extension of their wave functions.
This naturally leads to the question: How do inhomogeneous
magnetic field configurations alter the electronic structure of excited atoms ?

During the past decades many thorough investigations have been
performed on the behaviour and properties of excited (Rydberg-) atoms
in homogeneous magnetic fields (see the books and reviews \cite{Friedrich89,Ruder94,Friedrich97,
Schmelcher98,Schmelcher97}). Indeed investigations on atoms in strong magnetic
fields provided major contributions to a variety of different research areas such as
semiclassics of nonintegrable systems, 'quantum chaos', nonlinear dynamics, astrophysics
of magnetized stars and it elucidated and significantly advanced our understanding
of magnetized structures in general.

In contrast to the case of a homogeneous magnetic field there
exist no studies on the electronic structure of atoms in the
presence of inhomogeneous external fields: all investigations in
the literature on the behaviour of ultracold atoms in
inhomogeneous fields typically treat the atom as a point particle
whose magnetic moment couples either adiabatically \cite{Folman02}
or nonadiabatically
\cite{Bergeman89,Berg96,Burke96,Sukumar97,Hinds00,Potvliege01} to
the external field. This holds with the exception of two very
recent works \cite{Lesanovsky04_1, Lesanovsky04_2} that consider
the electronic structure of atoms with a single active electron
subject to a three-dimensional quadrupole field. A variety of
interesting new phenomena have been observed there. The symmetries
of this system cause each energy level to be degenerate in the
presence of the field. Furthermore the intimate coupling of the
spin and spatial degrees of freedom leads to a complex spatial
distribution of the spin polarization of individual electronic
states. A remarkable property of the electronic states in the
3D-quadrupole trap is the fact that they possess a magnetic
field-induced permanent electric dipole moment whose size strongly
varies with the Rydberg state considered. Besides the
3D-quadrupole field there is another generic inhomogeneous
magnetic field configuration which is employed to trap atoms in
particular on the atom chip \cite{Folman02}. This is the so-called
side guide which is created by superimposing the magnetic field of
a current carrying wire with a homogeneous bias field oriented
perpendicular to the wire. The resulting magnetic guide can be
augmented to a Ioffe-Pritchard type 2D-trap by applying an
additional homogeneous bias field parallel to the wire. It is
exactly this configuration which is studied in the present work
i.e. we investigate the structure and properties of electronically
excited atoms in a magnetic guide. According to the effects
obtained for atoms in a 3D-quadrupole trap in Refs.
\cite{Lesanovsky04_1,Lesanovsky04_2} we expect also the atoms in a
side guide to exhibit interesting new features.

The paper is organized as follows: In Sec. \ref{sec:hamiltonian} we introduce the field configuration
generated by a so-called side guide. We specify our
approach which is particularly suited for ultra-cold atoms with a
single active electron and derive the corresponding Hamiltonian.
This Hamiltonian exhibits a wealth of both unitary and anti-unitary symmetries and constitutes
an uncommon large symmetry group which is analyzed in Sec.
\ref{sec:symmetries}. In particular these symmetries lead to a two-fold degeneracy of any energy level, similar to the
case of an atom in a 3D-quadrupole trap. A discussion of an arbitrary spin-$\frac{1}{2}$-systems in a field
configuration obeying certain symmetries are discussed. Section \ref{sec:eigenstates}
contains a discussion of the properties of the symmetry-adapted
electronic states. In Sec. \ref{sec:ioffe} the latter are studied
in case an additional homogeneous (Ioffe-)field is applied. The numerical methods being employed in
order to solve the stationary Schr\"odinger equation are briefly
outlined in Sec. \ref{sec:numerics}. A discussion of our results
is provided in Sec. \ref{sec:results}. We analyze the spectra for
a wide range of gradients. Furthermore we explore
properties of the electronic spin such as spin expectation values and
distributions of the spin polarization. Selection rules and dipole
strengths of electric dipole transitions are calculated. We close
with a discussion of the electronic structure in case
a homogeneous magnetic field is applied in addition to the field of the magnetic guide. Sec.
\ref{sec:outlook} contains the summary and outlook.

%
\section{The Field Configuration and the Hamiltonian}\label{sec:hamiltonian}
\begin{figure}[htb]\center
\includegraphics[angle=0,width=7cm]{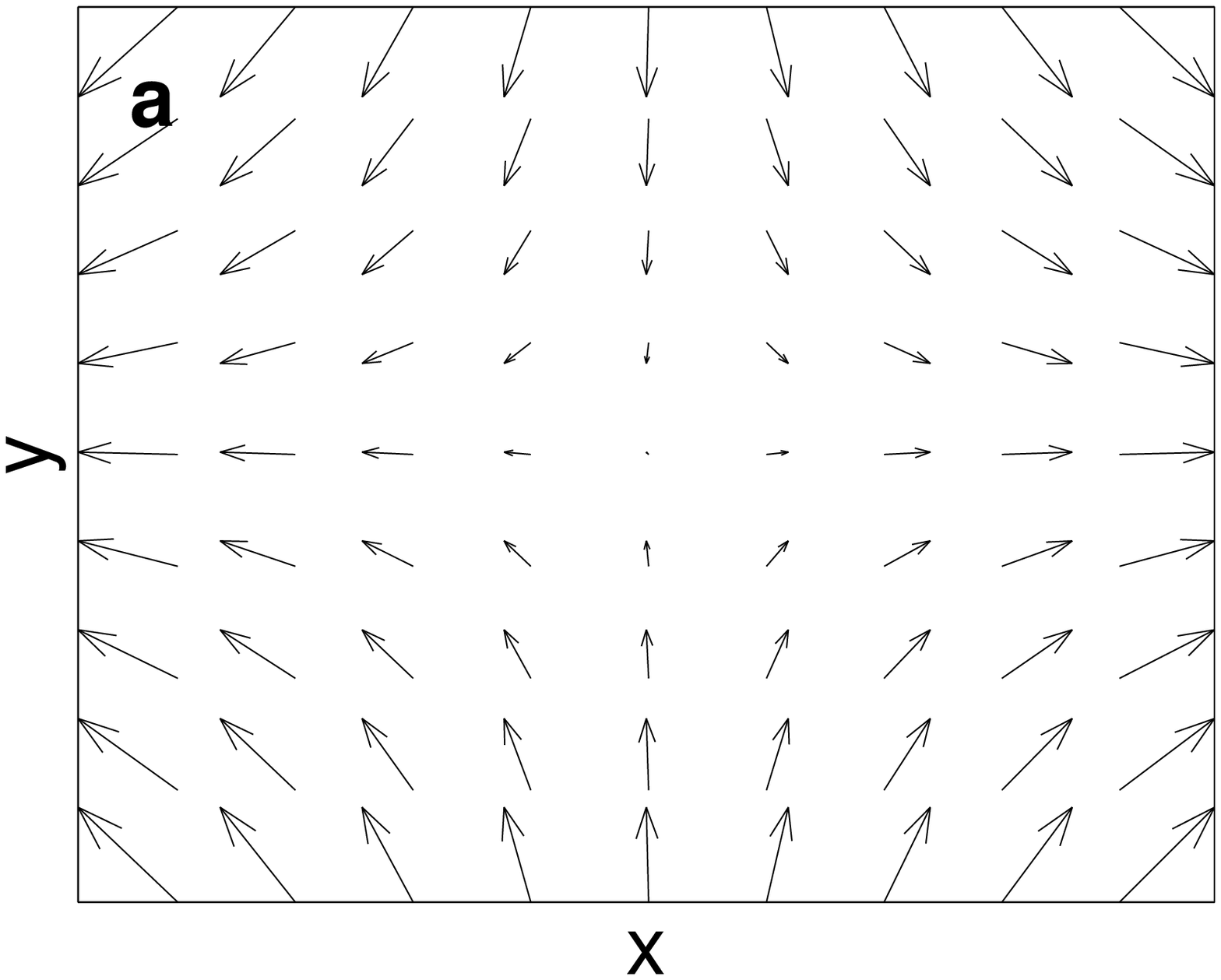}
\includegraphics[angle=0,width=7cm]{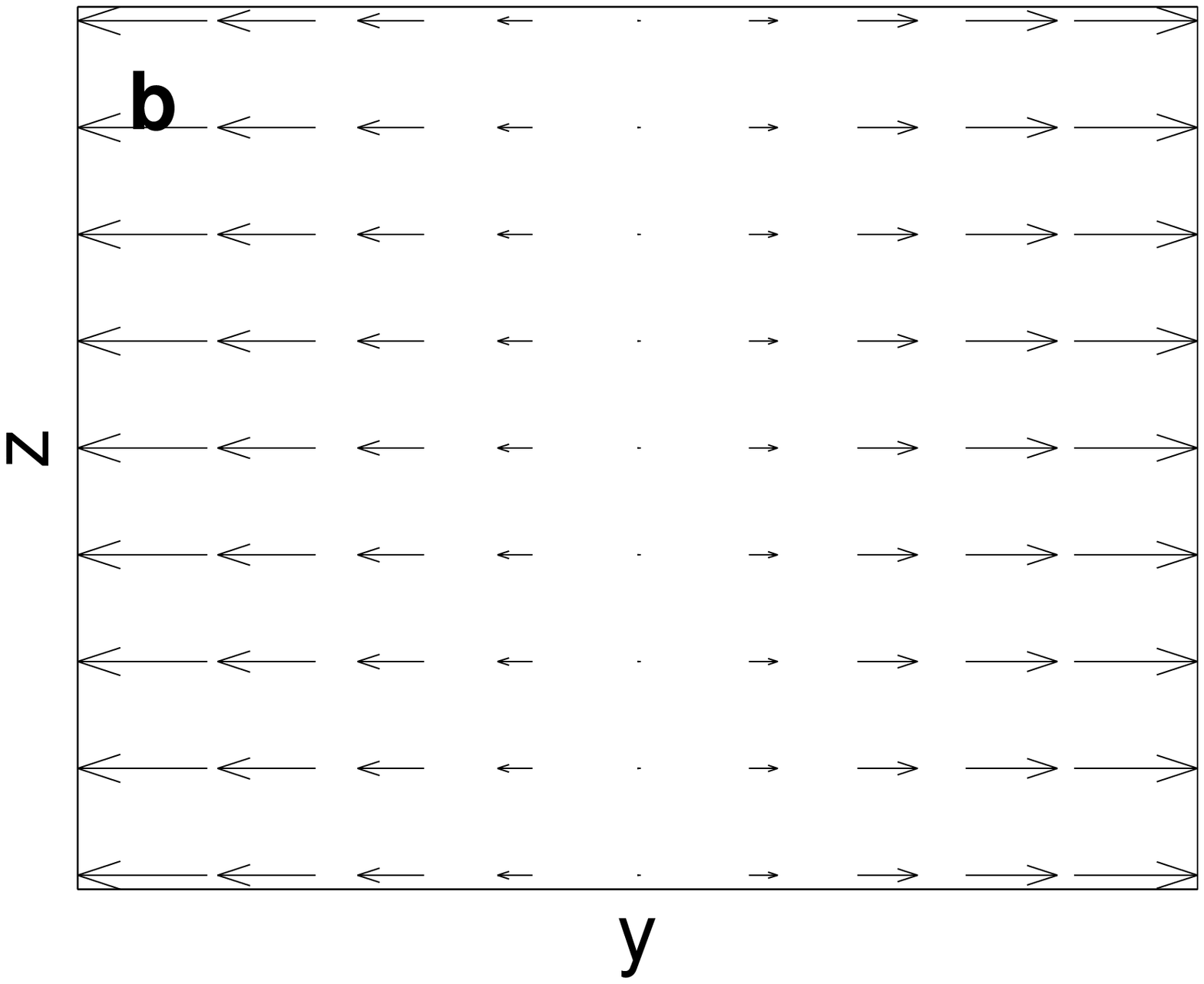}
\caption{Vectorial plots of the magnetic field
(\ref{eq:magnetic_field}). \textbf{a:} Intersection for
$z=0$. The quadrupolar shape of the field is clearly
recognized. \textbf{b:} Intersection for $x=0$
revealing the translational invariance with respect to the
$z$-coordinate.}\label{fig:magnetic_field}
\end{figure}

%
Alkali atoms are used throughout many experiments in ultra cold
atomic physics. Besides a single active electron they possess a
closed shell core and the total electronic spin is therefore
exclusively carried by the outer electron. We assume the motion of
this valence electron to take place in the Coulomb potential
of a single positive point charge. Since the focus of this work is
to understand fundamental features of electronically excited atoms
in a certain inhomogeneous magnetic field we do not account for
quantum defects which would require the consideration of
core-electron scattering processes. We also neglect relativistic
effects such as spin-orbit and hyperfine coupling. Both
interaction possess a $r^{-3}$-dependence with $r$ being the
distance between the outer electron and the nucleus. For (highly)
excited states their contributions can safely be neglected or, if
necessary accounted for by means of perturbation theory. Since we
focus on ultra cold atoms effects of the center of mass (c.m.)
motion on the electronic motion are neglected here. Specifically
we assume an infinitely heavy core (c.m.) located at the minimum
of the magnetic field. Employing the above approximations the
Hamiltonian describing the motion of the valence electron in the
presence of an external magnetic field reads
\begin{eqnarray}
H&=&\frac{1}{2
m_e}\left(\vec{p}+e\vec{A}\left(\vec{r}\right)\right)^2-\frac{e^2}{4\pi\epsilon_0
\left|\vec{r}\right|}+\frac{g_s\mu_B}{\hbar}\vec{S}\vec{B}\left(\vec{r}\right).\label{eq:hamiltonian_general}
\end{eqnarray}
The magnetic field is introduced via the minimal coupling
including the vector potential thereby providing the kinetic
energy in the presence of the field. The third term represents the
coupling between the spin of the electron and the external field.
A common configuration for the manipulation of neutral atoms is the
so-called magnetic side guide \cite{Folman02}. This particular
setup is generated by a current carrying wire whose 'circular'
magnetic field is superimposed by an external homogeneous bias
field perpendicular to the current flow. As a result the field
vanishes along a line parallel to the wire at a distance
$\rho_0=\frac{\mu_0 I}{2 \pi B}$ being completely determined by
the current $I$ and the homogeneous magnetic field strength $B$.
The Taylor expansion of the field around $\rho_0$ yields
\begin{eqnarray}
\vec{B}\approx\frac{B}{\rho_0}\left(%
\begin{array}{c}
  x \\
  -y \\
  0 \\
\end{array}%
\right)+\frac{B}{\sqrt{2}\rho_0^2}\left(%
\begin{array}{c}
  -x^2+2xy+y^2 \\
  x^2+2xy-y^2 \\
  0 \\
\end{array}%
\right)+\frac{2B}{3\rho_0^3}\left(%
\begin{array}{c}
  y\left(y^2-3x^2\right) \\
  -x\left(x^2-3y^2\right) \\
  0 \\
\end{array}%
\right).
\end{eqnarray}
These are the quadrupolar, hexapolar and octopolar expansion terms
of the field. Here we restrict ourselves to the linear term which
should provide a good approximation of the magnetic field
configuration as long as $\rho_0 \gg 1$. Thus we obtain the
expression
\begin{eqnarray}
\vec{B}=b\left(%
\begin{array}{c}
  x \\
  -y \\
  0 \\
\end{array}%
\right)\label{eq:magnetic_field}
\end{eqnarray}
Here $b$ is the magnetic field gradient determining the linear
growth of the field with increasing distance from the line of zero
field. Figure \ref{fig:magnetic_field} shows two vectorial plots
along cuts through the field.
The cut through the $x-y$-plane reveals the quadrupolar shape
of the field of the side guide whose translational invariance
along the $z$-axis can be easily observed in figure
\ref{fig:magnetic_field}b. A corresponding vector potential in the Coulomb
gauge is given by
\begin{eqnarray}
\vec{A}=b\left(%
\begin{array}{c}
  0 \\
  0 \\
  xy \\
\end{array}%
\right)\label{eq:vector_potential}
\end{eqnarray}
Inserting the expressions (\ref{eq:magnetic_field}) and
(\ref{eq:vector_potential}) into the Hamiltonian
(\ref{eq:hamiltonian_general}) thereby adopting atomic units
\footnote{$\hbar=1$, $m_e=1$, $a_0=1$: The magnetic gradient unit
then becomes $b=1a.u.=4.44181\cdot 10^{15} \frac{T}{m}$. The
magnetic field strength unit is $B=1a.u.=2.35051\cdot 10^{5} T$}
yields
\begin{eqnarray}
H&=&\frac{1}{2}\triangle-\frac{1}{\sqrt{x^2+y^2+z^2}}+b\,xyp_z+\frac{b^2}{2}\,x^2y^2+\frac{b}{2}\left(x\sigma_x-y\sigma_y\right)\label{eq:Hamiltonian_cart}
\end{eqnarray}
The first two terms of (\ref{eq:Hamiltonian_cart}) represent the
non-relativistic hydrogen atom. The third term which is linear with respect to
$b$ replaces the angular Zeeman term
\footnote{$\frac{B}{2}\left(xp_y-yp_x\right)$ where $B$ is the
field strength} which would occur in a homogeneous field. Here the
spatial coordinates in $x$ and $y$ couple with the momentum in $z$
direction. The successive diamagnetic term $\propto b^2$
represents an oscillator coupling term confining the electronic
motion in the $x$ and $y$ direction except for the axis exit
channels. This is reminiscent but also very different to the
situation in a homogeneous field, where the diamagnetic
interactions in the $x$ and $y$ direction separate and represent
pure harmonic oscillators. Finally the fifth term represents the
coupling of the electronic spin to the spatial coordinates arises
from the interaction of its magnetic moment with the field. We
here encounter a linear dependence on the spatial coordinates and
the gradient $b$. This term prevents the factorization of the
motions in coordinate space and spin space. Finally one should
note that the only explicit dependence on the coordinate $z$ is
due to the Coulomb term. Without this rotationally invariant
interaction the system would be invariant under translations with
respect to the $z$-coordinate.

Performing the canonical scaling transformation
$\bar{x}=b^\frac{1}{3}x$ and $\bar{p}=b^{-\frac{1}{3}}p$ the
Hamiltonian (\ref{eq:Hamiltonian_cart}) becomes
\begin{eqnarray}
 H=b^{-\frac{2}{3}}\bar{H}=\frac{1}{2}\vec{{p}}^{\,2}-\frac{\bar{Z}}{\sqrt{{x}^2+{y}^2+{z}^2}}
 +{x}{y}{p}_z+\frac{1}{2}\,{x}^2{y}^2
 +\frac{1}{2}\left({x}\sigma_x-{y}\sigma_y\right)\label{eq:Hamiltonian_cart_rescaled}
\end{eqnarray}
with $\bar{Z}=b^{-\frac{1}{3}}$ and where we have for simplicity
omitted the bar on top of the phase space variables. This shows us
that employing a scaled energy (scaled Hamiltonian) the only free
parameter is the scaled Coulomb coupling strength $\bar{Z}$ that
depends on the field gradient. The scaled Hamiltonian describes
the motion of an electron in the Coulomb-field of a charge
$\bar{Z}$ and the field with gradient $1$. If $b\rightarrow\infty$
the Coulomb term vanishes since $\bar{Z}\rightarrow 0$. In this
limit the energy level spacing is expected to scale according to
$b^{\frac{2}{3}}$.
%
\section{Symmetries and degeneracies in Spin-$\frac{1}{2}$-systems } \label{sec:symmetries}
%

In this section we analyze the structure of the Hamiltonian
(\ref{eq:Hamiltonian_cart}) in detail. After studying its
symmetries we discuss how these symmetries affect the excitation
spectrum. As a result of a tedious and elaborate analysis of the
Hamiltonian (\ref{eq:Hamiltonian_cart}) we found 15 distinct
symmetry operations leaving it invariant. A complete list is
provided in table \ref{tbl:symmetries}.
\begin{table}[htb]\center
\begin{tabular}{|c|c|c|c|}

\hline $\Sigma_x=\sigma_xP_yP_z$& $\Sigma_y=P_x\sigma_yP_z$& $\Sigma_z=P_xP_y\sigma_z$& \\

\hline $I_{xy}S_1^*$& $P_yP_zI_{xy}S_2$& $P_xP_yI_{xy}S_1$&$P_xP_zI_{xy}S_2^*$ \\

\hline\hline $T\sigma_xP_z$& $TP_xP_yP_z\sigma_y$& $TP_x\sigma_z$& $TP_y$\\

\hline $TP_yI_{xy}S_1^*$&$TP_zI_{xy}S_2$&
$TP_xI_{xy}S_1$&$TP_xP_yP_zI_{xy}S_2^*$
\\\hline

\end{tabular}\caption{Symmetry operations of the Hamiltonian
(\ref{eq:Hamiltonian_cart}). Top part: unitary symmetries. Bottom
part: anti-unitary symmetries.}\label{tbl:symmetries}
\end{table}
Each symmetry is composed of a number of elementary operations
which are shown in table \ref{tbl:operations}.
\begin{table}[htb]\center
\begin{tabular}{|c|c|c|}
\hline Operator& Operation& Designation\\

\hline\hline $P_{x_i}$& $x_i\rightarrow -x_i$& $x_i$-parity\\

\hline $T$& $A \rightarrow A^*$& conventional time reversal\\

\hline $\sigma_x$ &$\sigma_y\rightarrow
-\sigma_y$\quad$\sigma_z\rightarrow
-\sigma_z$& Pauli spin matrix x\\

\hline $\sigma_y$ &$\sigma_x\rightarrow
-\sigma_x$\quad$\sigma_z\rightarrow
-\sigma_z$& Pauli spin matrix y\\

\hline $\sigma_z$ &$\sigma_x\rightarrow
-\sigma_x$\quad$\sigma_y\rightarrow
-\sigma_y$& Pauli spin matrix z\\

\hline $I_{xy}$& $x \rightarrow y$\quad $y \rightarrow x$\quad\quad($\phi \rightarrow -\phi+\frac{\pi}{2}$)& coordinate exchange\\

\hline $S_1=\left(%
\begin{array}{cc}
  0 & 1 \\
  -i & 0 \\
\end{array}%
\right)$& $\sigma_x\rightarrow -\sigma_y$\quad$\sigma_y\rightarrow -\sigma_x$\quad $\sigma_z\rightarrow -\sigma_z$& \\

\hline $S_2=\left(%
\begin{array}{cc}
  -i & 0 \\
  0 & 1 \\
\end{array}%
\right)$& $\sigma_x\rightarrow -\sigma_y$\quad$\sigma_y\rightarrow
\sigma_x$\quad $\sigma_z\rightarrow \sigma_z$&
\\\hline
\end{tabular}\caption{Set of discrete operations out of which
 all symmetry operations of the Hamiltonian (\ref{eq:Hamiltonian_cart}) can be composed.
 Note that $S_1$ and $S_2$ are given in a basis where $\sigma_z$ is diagonal.}\label{tbl:operations}
\end{table}
All symmetry operations are either unitary or anti-unitary. The
anti-unitary ones involve the conventional time reversal operator
$T$. In spite of its simplicity our system therefore possesses a
wealth of symmetry properties. The algebra of the underlying
symmetry group possesses a complicated structure some features of
which will be discussed in the following. The operators
$\Sigma_x$, $\Sigma_y$ and $\Sigma_z$ generate a sub-group obeying
the algebra
$\left[\Sigma_i,\Sigma_j\right]=2i\,\epsilon_{ijk}\Sigma_k$
reminiscent of angular momentum operators. We have $\Sigma_i^2=1$.
Interestingly these quantities act on both real and spin space. A
deeper look into the representation theory of our group reveals a
two-fold degeneracy of any energy level similar to those we
encountered in our investigations of atoms in a three-dimensional
quadrupole trap \cite{Lesanovsky04_1,Lesanovsky04_2}.

Alternatively this degeneracy can also be established as follows. The
operations $\Sigma_z$ and $T\sigma_x P_z$ obey
$\left\{\Sigma_z,T\sigma_x P_z\right\}=0$. Let
$\left|E,\pi\right>$ be an energy eigenstate and at the same time
an eigenstate of $\Sigma_z$ with
\begin{eqnarray}
\Sigma_z\left|E,\pi\right> &=& \pi\left|E,\pi\right>
\end{eqnarray}
and $\pi=\pm 1$. Employing the above anti-commutator one obtains
\begin{eqnarray}
\Sigma_zT\sigma_x P_z\left|E,\pi\right> = -T\sigma_x
P_z\Sigma_z\left|E,\pi\right>= -\pi T\sigma_x
P_z\left|E,\pi\right>
\end{eqnarray}
The state $T\sigma_x P_z\left|E,\pi\right>$ can be identified with
$\left|E,-\pi\right>$. Hence, as long as $\pi\neq 0$
\footnote{Since $\Sigma_z$ is a unitary operator the case $\pi=0$
cannot occur.} there is always an orthogonal pair of states
possessing the same energy namely $\left|E,\pi\right>$ and
$\left|E,-\pi\right>$. We have to emphasize that there occur no
further degeneracies in the system. In principle one could think
of performing the above calculation repeatedly but now
substituting $T\sigma_x P_z$ with any operator listed in table
\ref{tbl:symmetries} which anti-commutes with $\Sigma_z$. It turns
out that all of the resulting states generated by this scheme are
either superpositions of $\left|E,\pi\right>$ and
$\left|E,-\pi\right>$ or differ only by a phase factor from one of
these states.

Out of the 15 symmetry operations one can choose several sets of
commuting operators. For the following investigation we choose the
set $H$, $\Sigma_z$, $P_yP_z I_{xy} S_2$. The combination of
$\Sigma_z$ and $P_yP_z I_{xy} S_2$ leads to the additional
commuting operator $P_x P_z I_{xy} S_2^*$. We have found the
properties:
\begin{eqnarray}
  \left(P_yP_z I_{xy} S_2\right)^2&=&\left(P_x P_z I_{xy}
S_2^*\right)^2=-\Sigma_z\label{eq:PPIS2}\\
\left(\Sigma_z\right)^2&=&\left(P_yP_z I_{xy}
S_2\right)^4=\left(P_x P_z I_{xy} S_2^*\right)^4=1.
\end{eqnarray}
For completeness we provide here the general embedding of the
above-derived degeneracies due to symmetries. Let us assume we
have a general spin-$\frac{1}{2}$-systems with the following
accompanying properties:
\begin{enumerate}
  \item{There are two operators $A$ and $B$ commuting with the
  underlying Hamiltonian:
  $\left[H,A\right]=\left[H,B\right]=0$.}
  \item{$A$ and $B$ anti-commute: $\left\{A,B\right\}=0$.}
  \item{$A$ is a Hermitian operator. $B$ is an (anti-)unitary operator which can be written as a
product $B=RS$ where $R$ and $S$ exclusively act on the real space
and the spin space, respectively.}
  \item{The operator $S$ is trace-less: $\text{Tr}\, S=0$.}
\end{enumerate}
If these conditions are fullfilled any state is doubly degenerate.
This is seen as follows. Property 4 immediately leads to $\text{Tr}\, B=0$. Hence, we find
the nonzero eigenvalues of $B$ to appear pairwise with opposite signs. If
now $\left|E,b\right>$ is an eigenstate of $B$ and at the same
time an energy-eigenstate property 2 implies that
\begin{eqnarray}
  BA\left|E,b\right>=-AB\left|E,b\right>=-bA\left|E,b\right>=-b\left|E,-b\right>.
\end{eqnarray}
Hence, $\left|E,b\right>$ and $A\left|E,b\right>=\left|E,-b\right>$
are two degenerate energy-eigenstates of the system.

In the present case the two anti-commuting operators are
$\Sigma_z$ and $T\sigma_xP_z$. For the case of an atom in a
three-dimensional quadrupole field we have $A=J_z$ and
$B=T\sigma_x P_z$. In a homogeneous magnetic field the
remaining symmetries constitute an Abelian symmetry group leading
to exclusively one dimensional irreducible representations i.e. no
degeneracies occur. Finally we remark that the reader can find in
ref. \cite{Haake01} a discussion of degeneracies in
spin-$\frac{1}{2}$-systems based on the properties of
time-reversal operators.
%
\section{$\Sigma_z$-, $P_yP_zI_{xy}S_2$- and $T\sigma_x P_z$-eigenstates
}\label{sec:eigenstates}
%

The operator $P_yP_zI_{xy}S_2$ obeys the eigenvalue relation
\begin{eqnarray}
  P_yP_z I_{xy} S_2\left|\kappa\right>=\kappa\left|\kappa\right>.
\end{eqnarray}
Since
\begin{eqnarray}
\left|\kappa\right>=\left(P_yP_z I_{xy}
S_2\right)^4\left|\kappa\right>=\kappa^4\left|\kappa\right>
\end{eqnarray}
the eigenvalue $\kappa$ can adopt the four values $\pm 1$ and $\pm
i$. The reader should note that $P_yP_z I_{xy} S_2$ is a unitary
but non-Hermitian operator. We therefore encounter complex
eigenvalues. If we apply $\Sigma_z$ to the states
$\left|\kappa\right>$ we find by exploiting equation
(\ref{eq:PPIS2})
\begin{eqnarray}
\Sigma_z\left|\kappa=\pm i\right>&=&\left|\kappa=\pm i\right> \label{eq:sigma_TPIS_eigenrelation_1}\\
\Sigma_z\left|\kappa=\pm 1\right>&=&-\left|\kappa=\pm 1\right>
\label{eq:sigma_TPIS_eigenrelation_2}.
\end{eqnarray}
By using the relation $\left(T\sigma_x P_z\right)\left(P_yP_z
I_{xy} S_2\right)- i\left(P_yP_z I_{xy}S_2\right)\left(T\sigma_x
P_z\right)=0$ one finds the degenerate pairs of states in the
$P_yP_z I_{xy}S_2$-subspaces:
$\left|E,+1\right>$,$\left|E,-i\right>$ and
$\left|E,-1\right>$,$\left|E,+i\right>$. Since non-Hermitian
operators do not represent physical observables only the quantum
number $\pi$ should be of direct relevance for the experimental
observation.

We now derive the expectation value of an observable $Y$ in an
eigenstate of $\Sigma_z$. Assume we have
$\left\{Y,\Sigma_z\right\}=0$ and hence
\begin{eqnarray}
  \left<E,\pi\right|Y \Sigma_z\left|E,\pi\right>&=&-\left<E,\pi\right|\Sigma_z
  Y\left|E,\pi\right>\\
  \pi\left<E,\pi\right|Y\left|E,\pi\right>&=&-\pi\left<E,\pi\right|Y\left|E,\pi\right>.
\end{eqnarray}
This immediately leads to the result
\begin{eqnarray}
  \left<E,\pi\right|Y\left|E,\pi\right>=0.\label{eq:Sigma_z_exp_value}
\end{eqnarray}
The same arguments hold for an observable $Z$ obeying
$\left\{Z,P_yP_z I_{xy} S_2\right\}=0$ in which case we obtain
\begin{eqnarray}
  \left<E,\kappa\right|Z\left|E,\kappa\right>=0.\label{eq:TPIS_exp_value}
\end{eqnarray}

In the preceding section we showed the degeneracy of the states
$\left|E,\pi\right>$ and $T\sigma_x P_z\left|E,\pi\right>$. By
superimposing these two states eigenstates of the operator
$T\sigma_x P_z$ can be constructed:
\begin{eqnarray}
\left|E,\pm\right>^{T\sigma_x
P_z}=\frac{1}{\sqrt{2}}\left[\left|E,\pi\right>\pm T\sigma_x
P_z\,\left|E,\pi\right>\right].\label{eq:TOP-state}
\end{eqnarray}
The corresponding eigenvalue relation is
\begin{eqnarray}
T\sigma_x P_z\,\left|E,\pm\right>=\pm\left|E,\pm\right>.
\end{eqnarray}

%
\section{Additional Homogeneous Field in $z$-Direction (Ioffe Field)}\label{sec:ioffe}
%
The application of an additional homogeneous magnetic field along
the $z$-direction (Ioffe field) has a dramatic impact on the
properties of the system. In particular the symmetry properties
are affected. The Hamiltonian becomes
\begin{eqnarray}
H_I&=&-\frac{1}{2}\triangle_{x,y,z}-\frac{1}{\sqrt{x^2+y^2+z^2}}+b\,xyp_z+\frac{b^2}{2}\,x^2y^2+\frac{b}{2}\left(x\sigma_x-y\sigma_y\right)\nonumber\\
&&+\frac{B_I}{2}\left(xp_y-yp_x\right)+\frac{B_I^2}{8}\left(x^2+y^2\right)+\frac{B_I}{2}\sigma_z
\label{eq:Hamiltonian_Ioffe_cart}
\end{eqnarray}
with $B_I$ being the field strength of the Ioffe field. Since both
the 2D-quadrupole (due to the side guide) and the magnetic field are perpendicular to each
other the homogeneous field terms can simply be added to the
Hamiltonian (\ref{eq:Hamiltonian_cart}). We find the well known
Zeeman as well as the diamagnetic oscillator term. The coupling of
the spin to the Ioffe field leads to a term being proportional to
$\sigma_z$. The symmetries of $H_I$ are listed in table
\ref{tbl:symmetries_Ioffe}.
\begin{table}[htb]\center
\begin{tabular}{|c|c|c|c|}

\hline $\Sigma_z$&$P_yP_zI_{xy}S_2$&$P_xP_zI_{xy}S_2^*$&\\

\hline\hline $TP_x\sigma_z$ & $TP_zI_{xy}S_2$&$TP_xP_yP_zI_{xy}S_2^*$ &$TP_y$\\

\hline

\end{tabular}\caption{Symmetries of the Hamiltonian
(\ref{eq:Hamiltonian_Ioffe_cart}), i.e. side guide with Ioffe
field. Top line: unitary symmetries. Bottom line: anti-unitary
symmetries.}\label{tbl:symmetries_Ioffe}
\end{table}
Due to the presence of the additional homogeneous field numerous
symmetries are lost (see table \ref{tbl:symmetries} for
comparison). The remaining operations form a non-Abelian algebra.
In contrast to the group operations listed in table
\ref{tbl:symmetries} there are no two anti-commuting operators.
Hence it is not possible to construct pairs of degenerate energy
eigenstates as discussed above. Thus, applying the Ioffe field
lifts the degeneracies occuring in the absence of it.
Even with a finite Ioffe field the operations $\Sigma_z$, $P_yP_z
I_{xy} S_2$ and $P_x P_z I_{xy} S_2^*$ together with $H_I$ form a
set of commuting operators.

%
\section{Numerical Treatment}\label{sec:numerics}
%

In order to obtain many eigenvalues and eigenfunctions of the
Hamiltonians (\ref{eq:Hamiltonian_cart}) and
(\ref{eq:Hamiltonian_Ioffe_cart}) particularly for highly excited
Rydberg states we adopt the linear variational principle. Here the
bound state solutions of the Schr\"odinger equation are expanded
in a finite set of square integrable basis functions. Determining
the expansion coefficients is equivalent to solving a generalized
eigenvalue problem in case of non-orthogonal basis functions. The
latter is done numerically by employing standard linear algebra
techniques and routines.

To accomplish the above we adopt spherical coordinates. The
Hamiltonian (\ref{eq:Hamiltonian_cart}) then becomes
\begin{eqnarray}
H&=&-\frac{1}{2}\triangle_{r,\theta,\phi}-\frac{1}{r}-ibr\sin\phi\cos\phi\left(\sin^2\theta\cos\theta\,
r\frac{\partial}{\partial r}-\sin^3\theta\frac{\partial}{\partial \theta}\right)\nonumber\\
&&+\frac{b^2}{2}r^4\sin^4\theta\sin^2\phi\cos^2\phi+\frac{b}{2}r\sin\theta\left(%
\begin{array}{cc}
  0 & e^{i\phi} \\
  e^{-i\phi} & 0 \\
\end{array}%
\right).\label{eq:Hamiltonian_spher}
\end{eqnarray}
With an additional Ioffe applied we have to consider the
Hamiltonian (\ref{eq:Hamiltonian_Ioffe_cart}) which reads in
spherical coordinates
\begin{eqnarray}
  H_I=H-i\frac{B_I}{2}\frac{\partial}{\partial\phi}+\frac{B_I^2}{8}r^2\sin^2\theta+\frac{B_I}{2}\sigma_z.
\end{eqnarray}
We utilize a Sturmian basis set of the form
\begin{eqnarray}
\left|n,l,m,s\right>=R_n^{(\zeta,k)}(r)\,Y_l^m(\theta,\phi)\left|s\right>.\label{eq:basis_functions}
\end{eqnarray}
These functions form a complete set in real and spin space but are
not orthogonal. The angular part is covered by the well-known
spherical harmonics $Y_l^m(\theta,\phi)$ whereas the two spinor
components are addressed by the spin-orbitals
$\left|s\right>=\left|\uparrow\right>$ or
$\left|\downarrow\right>$. For the radial part we employ
\begin{eqnarray}
R_{n}^{(\zeta,k)}(r)=\sqrt{\frac{n!}{(n+2k)!}}e^{-\frac{\zeta
r}{2}}(\zeta r)^{k}L_n^{2k}(\zeta r).\label{eq:radial_part}
\end{eqnarray}
with $L_n^{2k}(r)$ being the associated Laguerre polynomials. The
parameters $k$ and $\zeta$ can be adapted in order to gain an
optimal convergence behavior in any spectral region. In particular
the non-linear variational parameter $\zeta$ has to adapted such
that it corresponds to the inverse of the characteristic length
scale of the desired wavefunctions. Similar basis sets have been
employed previously by several other authors
\cite{Clark80,Clark82,Wunner86}.

The general expansion of an energy eigenstate $\left|E\right>$ in
a finite set of basis functions (\ref{eq:basis_functions}) reads
\begin{eqnarray}
\left|E\right>=\sum_{nlms}c_{nlms}\left|n,l,m,s\right>.
\end{eqnarray}
From our knowledge of the symmetries of the system we can further
specify the appearance of the expansion. In section
\ref{sec:symmetries} we chose $H$, $\Sigma_z$ and $P_yP_z I_{xy}
S_2$ to be the set of commuting operators whose eigenfunctions we
want to construct. We now demand $\left|E\right>$ to be an
eigenstate of $P_yP_z I_{xy} S_2$. Exploiting the relations
\begin{eqnarray}
  P_yP_zI_{xy}S_2\,Y_l^m\left|\uparrow\right>&=&-ie^{-i\frac{\pi}{2}m}(-1)^l\,Y_l^m\left|\uparrow\right>\\
P_yP_zI_{xy}S_2\,Y_l^m\left|\downarrow\right>&=&e^{-i\frac{\pi}{2}m}(-1)^l\,Y_l^m\left|\downarrow\right>.
\end{eqnarray}
we construct the following expansions for the four
$\kappa$-subspaces
\begin{eqnarray}
\left|E,+1\right>&=&\sum_{nlm}\left[R_n(a_{nlm}Y_{2l+1}^{4m+1}+b_{nlm}Y_{2l}^{4m+3})\left|\uparrow\right>+\bar{R}_n(c_{nlm}Y_{2l+1}^{4m+2}+d_{nlm}Y_{2l}^{4m+4})\left|\downarrow\right>\right]\label{eq:TPIS_func_1}\\
\left|E,-1\right>&=&\sum_{nlm}\left[R_n(a_{nlm}Y_{2l}^{4m+1}+b_{nlm}Y_{2l+1}^{4m+3})\left|\uparrow\right>+\bar{R}_n(c_{nlm}Y_{2l}^{4m+2}+d_{nlm}Y_{2l+1}^{4m+4})\left|\downarrow\right>\right]\\
\left|E,+i\right>&=&\sum_{nlm}\left[R_n(a_{nlm}Y_{2l}^{4m+2}+b_{nlm}Y_{2l+1}^{4m+4})\left|\uparrow\right>+\bar{R}_n(c_{nlm}Y_{2l+1}^{4m+1}+d_{nlm}Y_{2l}^{4m+3})\left|\downarrow\right>\right]\\
\left|E,-i\right>&=&\sum_{nlm}\left[R_n(a_{nlm}Y_{2l+1}^{4m+2}+b_{nlm}Y_{2l}^{4m+4})\left|\uparrow\right>+\bar{R}_n(c_{nlm}Y_{2l}^{4m+1}+d_{nlm}Y_{2l+1}^{4m+3})\left|\downarrow\right>\right]\label{eq:TPIS_func_4}
\end{eqnarray}
The eigenfunctions (\ref{eq:TPIS_func_1}-\ref{eq:TPIS_func_4}) are
automatically also eigenfunctions to $\Sigma_z$ (see eq.
(\ref{eq:sigma_TPIS_eigenrelation_1}) and
(\ref{eq:sigma_TPIS_eigenrelation_2})). Due to the structure of
the spherical harmonics $Y_l^m$ one has to ensure that
$\left|m\right|\leq l$. In our calculations the sums run over all
valid combinations of $n\leq N$, $l\leq L$ and $m\leq M$ where the
maximum indices $N$, $L$ and $M$ can be fixed individually. The
expansion becomes exact if $M,N,L \rightarrow \infty$.

Performing the linear variational principle with one of the above
expansions leads to a generalized eigenvalue problem
$\mathbf{H}\vec{v}=E\mathbf{S}\vec{v}$, where $\mathbf{H}$ and
$\mathbf{S}$ are the corresponding matrix representation of the
Hamiltonian (\ref{eq:Hamiltonian_spher}) and the overlap matrix,
respectively:
\begin{eqnarray}
  \mathbf{H}=\left<E,\kappa\right|H\left|E,\kappa\right> &\qquad&
  \mathbf{S}=\left<E,\kappa\mid E,\kappa\right>.
\end{eqnarray}
The vector $\vec{v}$ contains the expansion coefficients $a_{nlm}$,
$b_{nlm}$, $c_{nlm}$ and $d_{nlm}$.

Due to the particular choice of the basis functions
(\ref{eq:basis_functions}) the matrices $\mathbf{H}$ and
$\mathbf{S}$ become extremely sparse occupied ($\mathbf{S}$ is a
penta-banded matrix). In order to solve the generalized eigenvalue
equation we utilize the so-called Arnoldi method together with the
shift-and-invert method. We adopt routines from the ARPACK
package. A more detailed description can be found in
\cite{Lesanovsky04_2}.

%
\section{Results and Discussion}\label{sec:results}
%
In this section we analyze our computational results i.e. the
eigenvalues and eigenfunctions obtained via the numerical approach described in the previous section.
We discuss the spectra and expectation
values of several observables as well as the properties of the
electronic spin. Furthermore selection rules for electric dipole transitions as well as their strengths
are derived. Results for the case of the additional presence of a homogeneous bias field
are presented as well.

\subsection{Spectral Properties}
\begin{figure}[htb]\center
\includegraphics[angle=0,width=7cm]{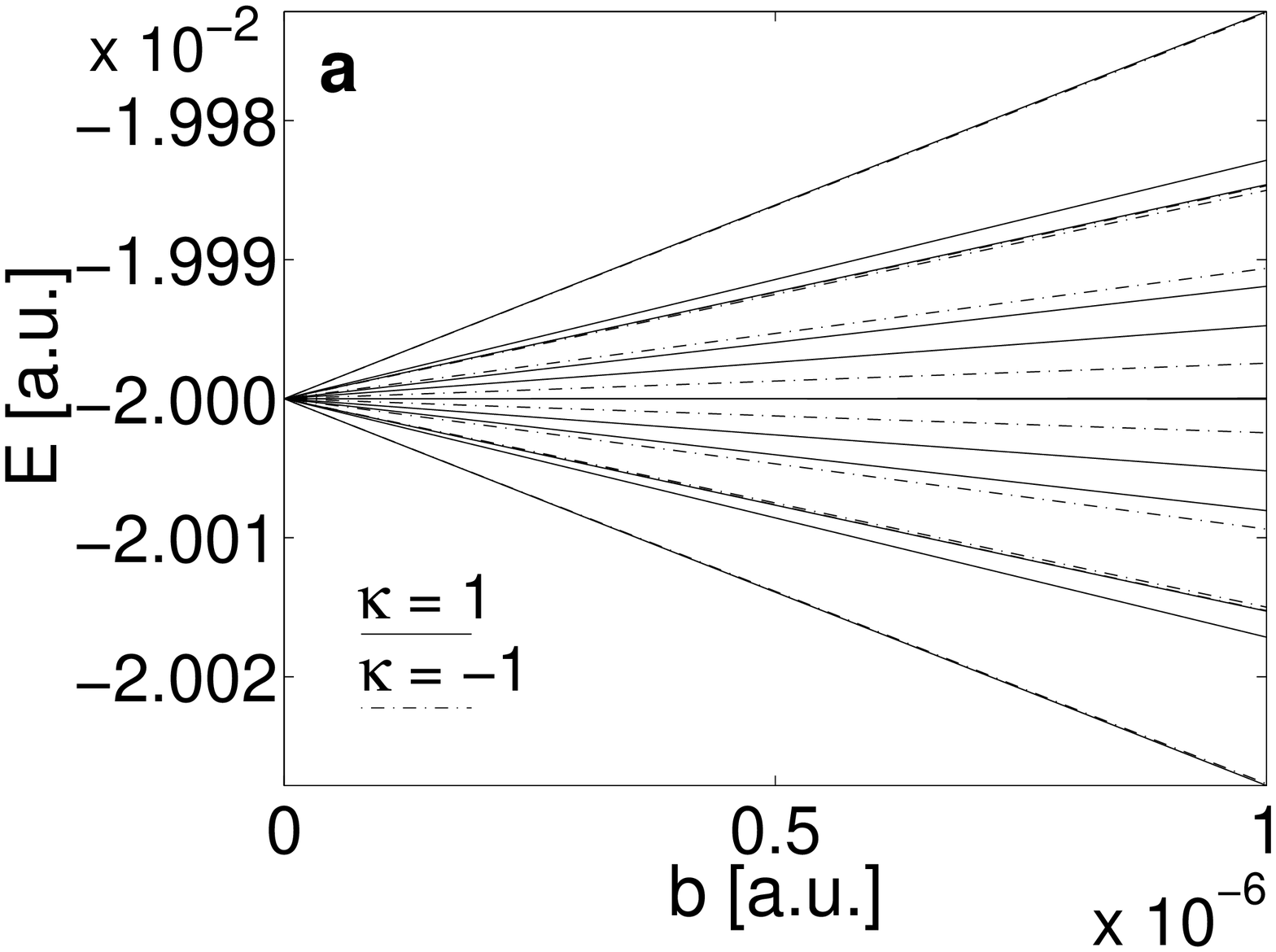}
\includegraphics[angle=0,width=7cm]{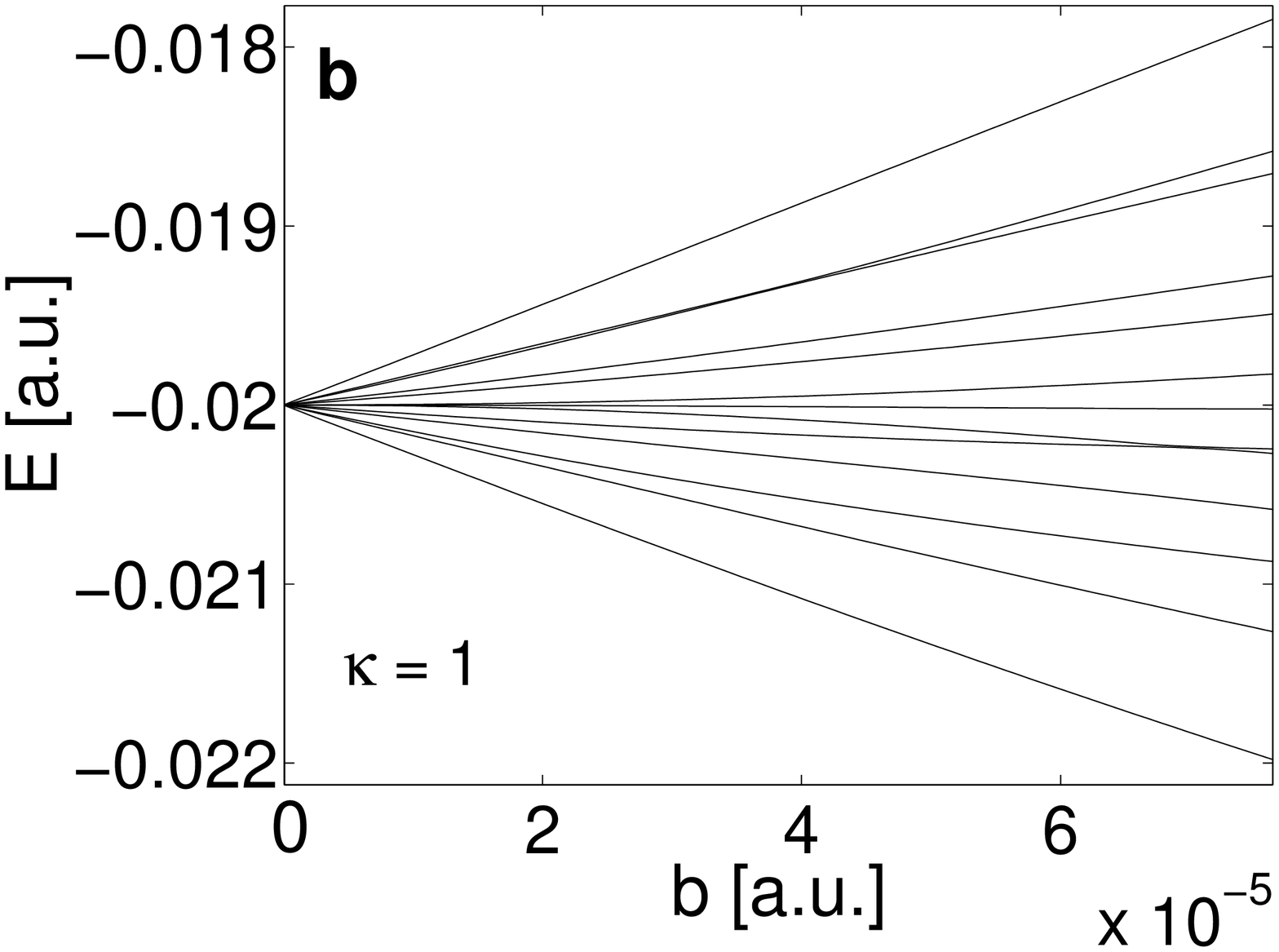}
\caption{\textbf{a:} Splitting of the energy levels belonging to
the $n=5$ multiplet ($\kappa=\pm 1$-subspace) with increasing
gradient. The level structure is dominated by the linear Zeeman
term. The splitting is linear and symmetric around the energy for
$b=0$. \textbf{b:} Intra $n$-manifold mixing of the $n=5$
multiplet in the $\kappa=1$-subspace. Due to the increasing
dominance of the diamagnetic term the level splitting becomes
non-linear.} \label{fig:n_5_splitting}
\end{figure}

With respect to the spectral behavior one can distinguish three
regimes: the weak, the intermediate and the strong gradient regime
each of which reveals individual characteristics. The appearance
of these regimes is not determined by the gradient and the degree
of excitation, i.e. energy, but by the scaled energy (see
discussion in Sec. \ref{sec:introduction}). For simplicity we will
refer to the gradient as the relevant quantity characterizing the different
regimes. All figures in this subsection show energy levels for
manifolds belonging to rather small values for $n$ (typically $(n=5\, -\, 7)$)
and for large gradients (we cover the range $b=10^{-7}-10^{-4}$) that are not accessible in
the laboratory. This was done for reasons of illustration: Our observations
and results equally hold for weaker gradients and higher $n$-manifolds
(gradients achievable for tight traps on atom chips are of the order of $b=10^{-8}$)
which however, due to the high level density, are less suited for a graphical presentation.
In the weak gradient regime the spectral behaviour is determined
by the linear Zeeman terms. Although the principal
quantum number $n$ is not a good quantum number any given level
can be assigned to a certain $n$-multiplet. The levels split
symmetrically around the zero-field-energy exhibiting the expected
linear dependence on $b$. In figure \ref{fig:n_5_splitting}a this
is exemplarily shown for the $n=5$-multiplet.

\begin{figure}[htb]\center
\includegraphics[angle=0,width=7cm]{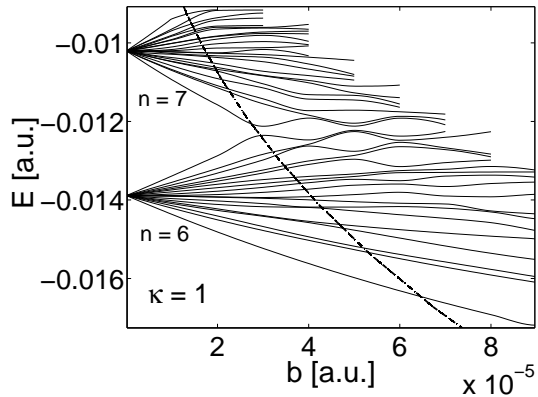}
\caption{Inter $n$-manifold mixing between the $n=6$- and
$n=7$-multiplet in the $\kappa=1$-subspace. The mixing threshold is
indicated by the dashed line. A large number of avoided crossings
occur.} \label{fig:inter_n_mixing}
\end{figure}

The intermediate regime is characterized by the occurence of intra
$n$-manifold mixing. Although neighboring $n$-manifolds are still
distinguishable the levels now aquire a a nonlinear $b$-dependence
which is due to the increasing importance of the diamagnetic term.
Sub-levels belonging to different angular momenta mix and thus
avoided level-crossings appear. The onset of this intermediate
regime scales according to $b\propto n^{-6}$. Figure
\ref{fig:n_5_splitting}b shows the regime of intermediate
gradients of the $n=5$-multiplet. Interestingly we observe here that
this nonlinear behaviour in the $l-$mixing regime is very weakly
pronounced for the atom in the side guide compared to an atom
in a homogeneous magnetic field \cite{Friedrich89}.
As we enter the strong gradient regime adjacent $n$-manifolds begin
to overlap. The spectra are strongly i.e. nonperturbatively influenced by the
diamagnetic term. Figure \ref{fig:inter_n_mixing} shows this inter
$n$-manifold mixing for the $n=6$- and $n=7$-multiplet where the
strong coupling leads to large avoided crossings. The mixing
threshold scales according to $b\propto n^{-\frac{11}{2}}$ (indicated by
the dashed line in figure \ref{fig:inter_n_mixing}).

\subsection{Properties of the Electronic Spin}

\subsubsection{$S_z$ expectation value}

In order to study the mutual influence of coordinate and spin
space let us investigate the properties of the electronic
spin. The $x$- and $y$-components of the spin operator obey
$\left\{\Sigma_z,S_x\right\}=\left\{\Sigma_z,S_y\right\}=0$. Hence
using (\ref{eq:Sigma_z_exp_value}) we arrive at
\begin{eqnarray}
  \left<S_x\right>=\left<S_y\right>=0.
\end{eqnarray}
Only the expectation value of $S_z$ is non-zero in general. This is
not obvious since the Hamiltonian (\ref{eq:Hamiltonian_spher}) does
not contain an explicit dependence on $S_z$.
\begin{figure}[htb]\center
\includegraphics[angle=0,width=7cm]{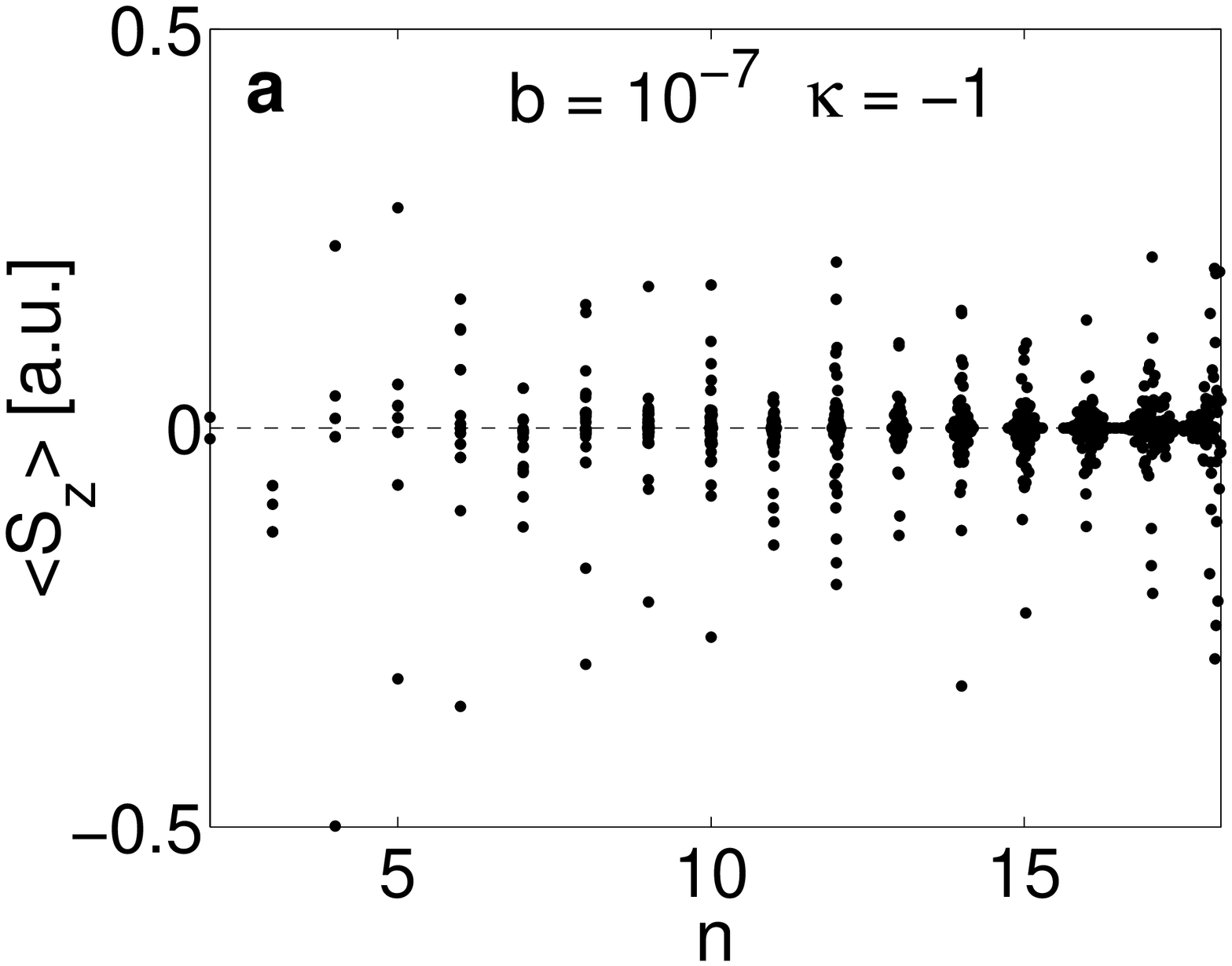}
\includegraphics[angle=0,width=7cm]{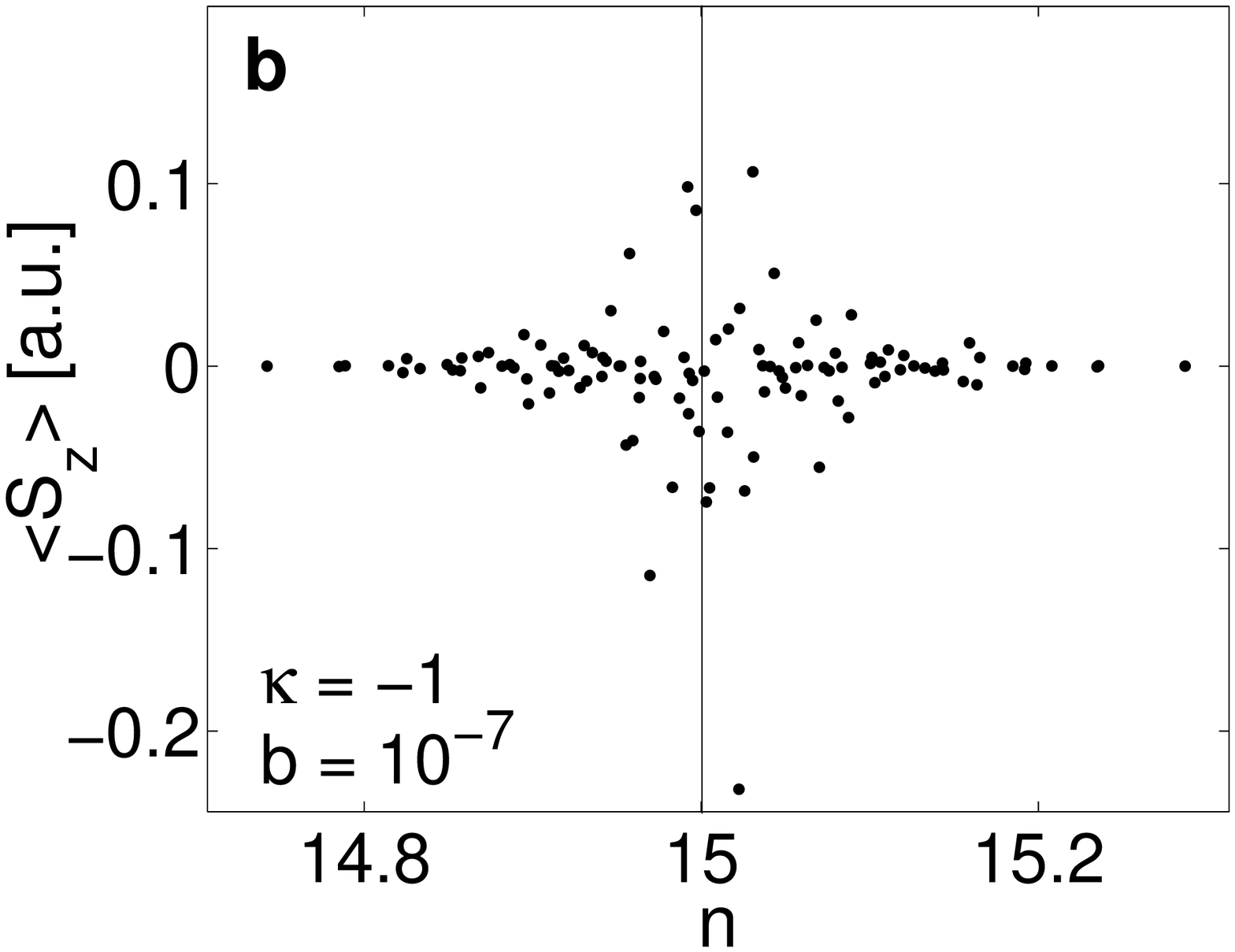}
\caption{\textbf{a:} Expectation value of the $z$-component of the
electronic spin operator for several excited states ($b=10^{-7}$).
\textbf{b:} Zoomed view of the $n=15$-multiplet. The magnitude of
$\left<S_z\right>$ decreases for states possessing a large energy
shift due to the external field. } \label{fig:sz_exp_value}
\end{figure}
Figure \ref{fig:sz_exp_value}a shows the expectation value
$\left<S_z\right>$ for several excited states as a function of the
principal quantum number $n$, which serves as an energetic label.
The expectation
values are arranged along vertical lines each of which belongs to
a certain $n$-multiplet. With increasing degree of excitation
these lines widen and begin to overlap as the inter $n$-mixing
regime is reached. A zoomed view of the $n=15$-multiplet is shown
in figure \ref{fig:sz_exp_value}b. We find states experiencing a
large energy shift due to the external field thereby possessing a small
$S_z$ expectation value. For the states shown in this figure
$\left<S_z\right>$ vanishes for  $n>15.2$ and $n<14.8$.

\subsubsection{Spatial Distributions of the Spin Polarization}
We now study the relative alignment of the electronic spin and the
magnetic field. For a two-component spinor
$\left|\Psi\right>=\left(\left|u\right>,\left|d\right>\right)^T$
we define
\begin{eqnarray}
  W_{S B}(\vec{r})&=&\frac{\left<\Psi\mid\vec{r}\right>\left<\vec{r}\right|\vec{S}\vec{B}\left|\vec{r}\right>\left<\vec{r}\mid\Psi\right>}{\left|\vec{S}\right|
  |\vec{B}|\left|\left<\Psi\mid\vec{r}\right>\right|^2}
  =\frac{\left<\Psi\mid\vec{r}\right>(\sigma_x \cos\phi-\sigma_y \sin\phi)\left<\vec{r}\mid\Psi\right>}{\left|\left<\vec{r}\mid u\right>\right|^2+\left|\left<\vec{r}\mid
  d\right>\right|^2}\nonumber\\
  &=&2\frac{\text{Re}\left[u^*(\vec{r})d(\vec{r})e^{i\phi}\right]}{\left|u(\vec{r})\right|^2+\left|d(\vec{r})\right|^2}
  =\left<\cos\gamma\right>\left(\vec{r}\right)\label{eq:SB_polarization}
\end{eqnarray}
$W_{S B}(\vec{r})$ describes the spatial distribution of the spin
polarization relative to the local magnetic field. $W_{S
B}(\vec{r})=1$ indicates the spin to be oriented parallel to the
field whereas we find it antiparallel aligned for $W_{S
B}(\vec{r})=-1$. According to (\ref{eq:SB_polarization})  $W_{S
B}(\vec{r})$ can be interpreted as the local expectation value of
the cosine of the angle $\gamma$ between $\vec{S}$ and $\vec{B}$. Since in a
homogenous field the projection of the spin onto the field
direction is conserved $W_{S B}(\vec{r})$ would be either $+1$ or
$-1$ throughout the whole space. In the field of the side guide, however,
we expect a much richer structure resulting from the coupling of
the coordinate and the spin degrees of freedom.
\begin{figure}[htb]\center
\includegraphics[angle=0,width=5.2cm]{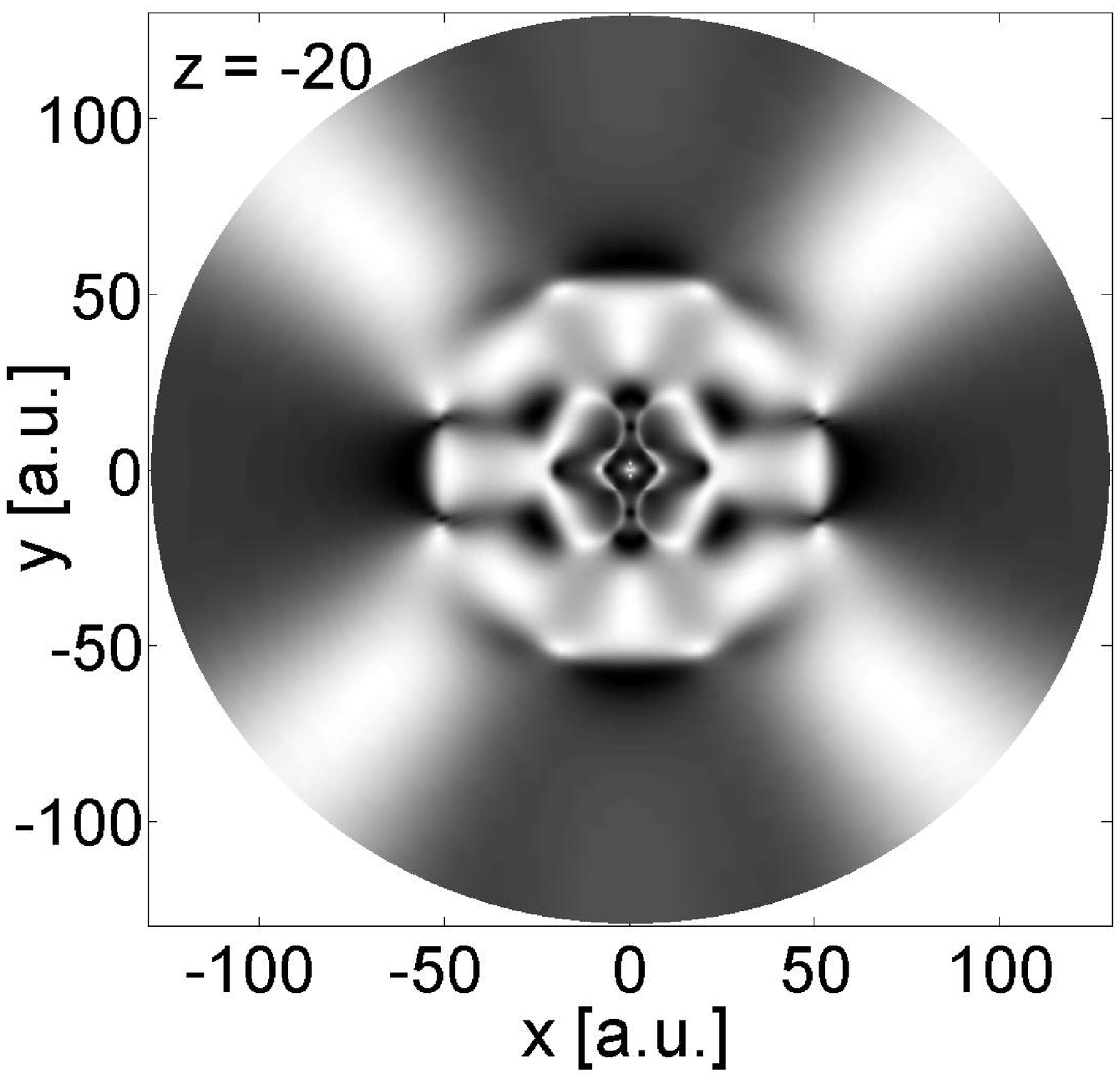}
\includegraphics[angle=0,width=5.2cm]{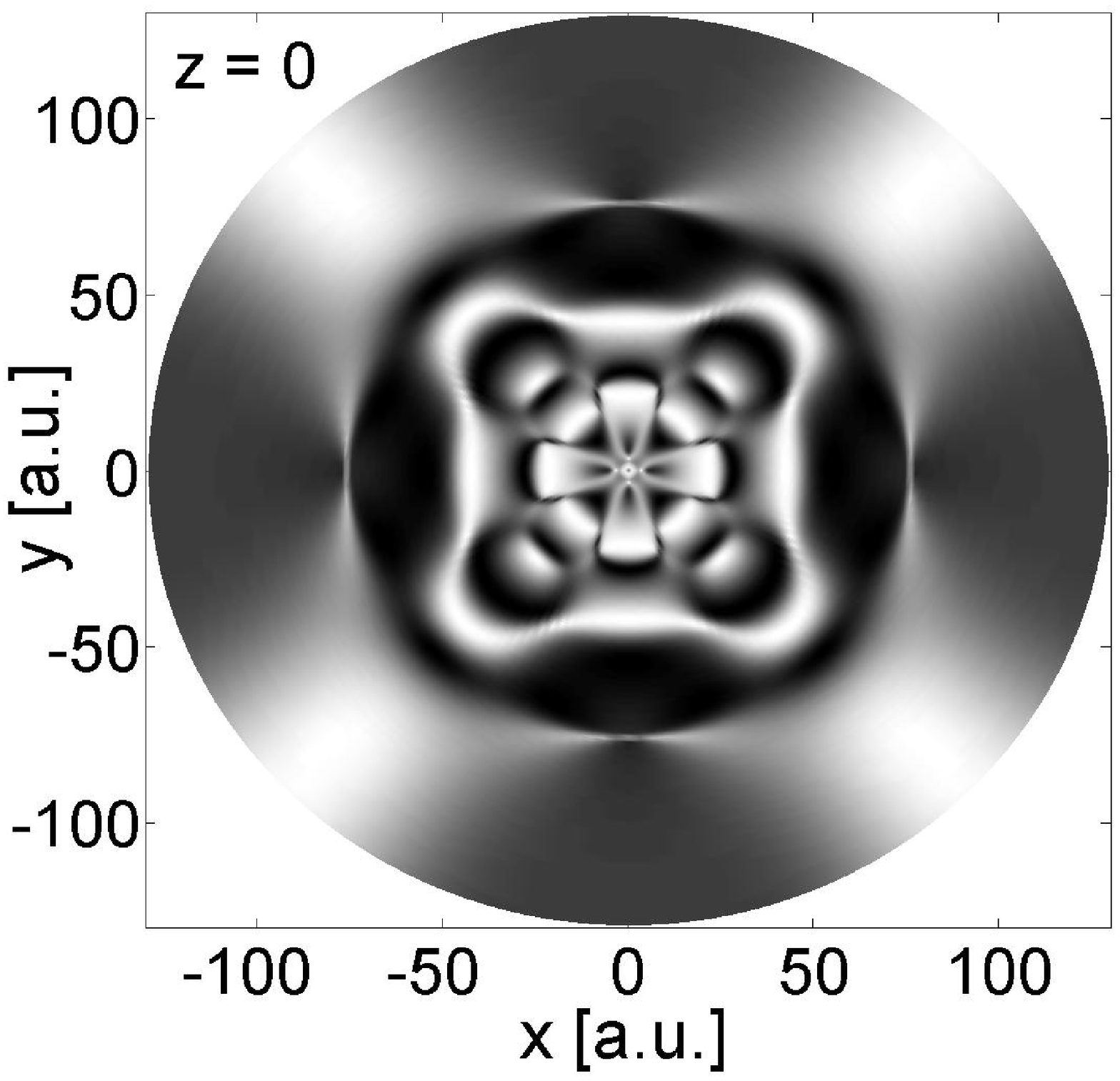}
\includegraphics[angle=0,width=5.2cm]{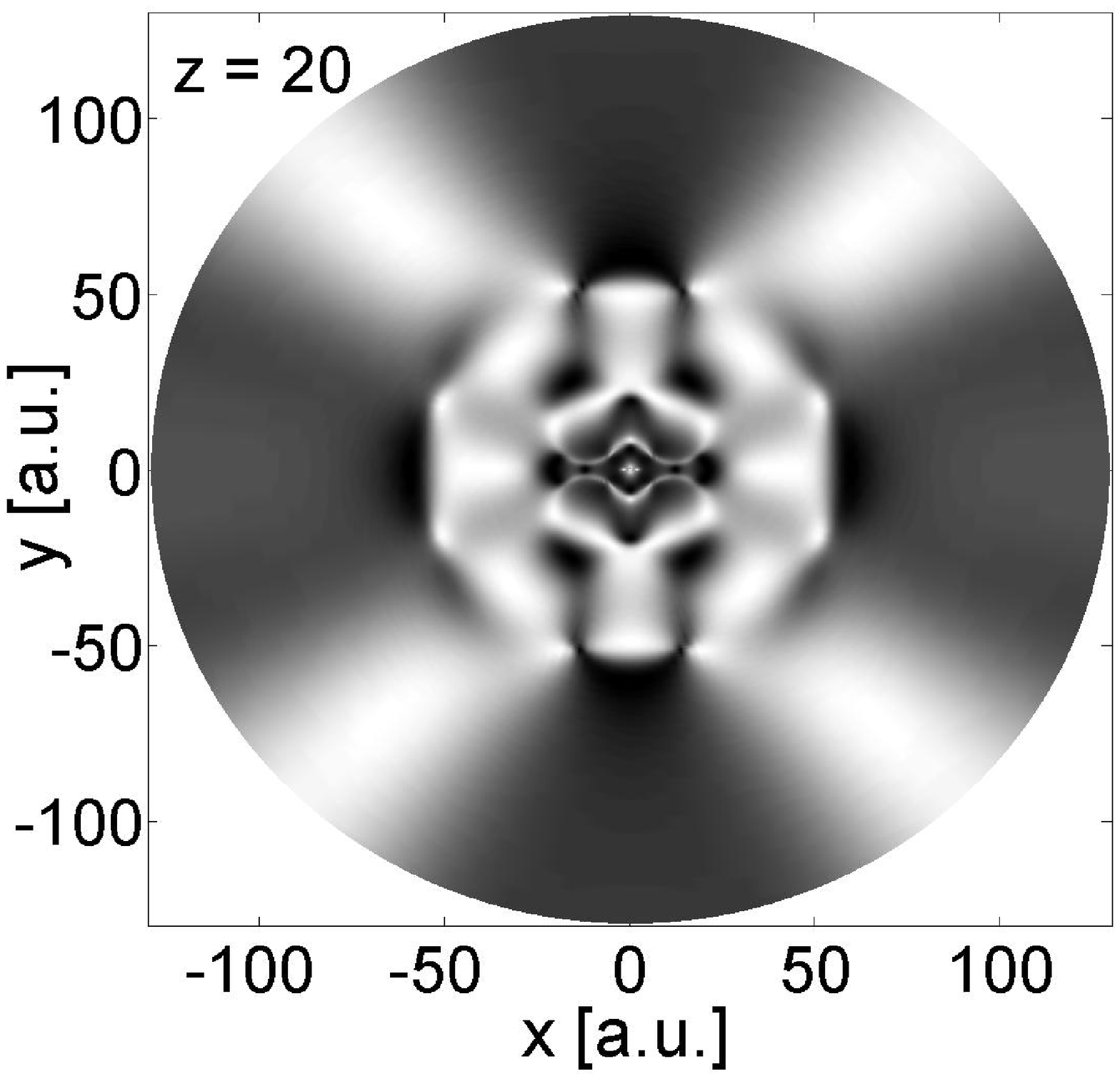}
\caption{Tomographic cuts through the spin polarization $W_{SB}$
of the $83$rd excited state. The state belongs to the $n=8$
multiplet within the $\kappa=1$-subspace ($b=10^{-7}$). The cuts
are made at $z=\pm 20$ and $z=0$. Positive and negative values are
indicated white and black, respectively. We observe a rich pattern
of different spin polarizations around the origin. From $r\approx
60$ on the nodal structure is replaced by a regular striped
pattern varying periodically with the azimuthal angle $\phi$.}
\label{fig:SB_density}
\end{figure}
Figure \ref{fig:SB_density} shows three tomographic cuts of a the
spin polarization $W_{SB}$ of the $83rd$ excited state in the
$\kappa=1$-subspace. In the vicinity of the coordinate center we
observe a large number of nodes.  From $\rho\approx 60$ on the
complex nodal structure is replaced by a smooth regular pattern
exhibiting a periodicity with respect to the azimuthal angle
$\phi$. Here $W_{S B}(\vec{r})$ becomes almost independent of the
$z$-coordinate. This feature seems to be induced mainly by the
magnetic interaction which is invariant under translations along
$z$. One identifies four sectors reminiscent of the quadrupolar
structure of the magnetic field of the guides. In the present case
we apparently have a anti-parallel alignment in the $x=0$- and
$y=0$-plane and a parallel one between these planes. The densities
are invariant under the operations $P_x P_y$ and $P_z I_{xy}$
which are equivalent to $\Sigma_z$ and $P_y P_z I_{xy} S_2$ when
acting on real and scalar quantities.

%
\subsection{Electric Dipole Transitions}
%
We now consider electromagnetic transitions between electronic
states in the framework of the dipole approximation. The
transition amplitude between the initial state $\left|i\right>$
and the final state $\left|f\right>$ is then given by the squared
modulus of the matrix element $\left<i\right| D \left|f\right>$.
In the length gauge $D$ takes the forms
$D_{\sigma^\pm}=\frac{1}{\sqrt{2}}\left(x\pm
iy\right)=\frac{1}{\sqrt{2}}r\sin\theta e^{\pm i \phi}$ and
$D_\pi=z=r\cos\theta$ for $\sigma^\pm$- and $\pi$-transitions,
respectively.

Exploiting the symmetry properties of the $P_y P_z I_{xy}
S_2$-eigenstates yields
\begin{eqnarray}
\left<E,\kappa\right|\left(P_y P_z I_{xy}S_2\right)^+ z \,P_y P_z
I_{xy} S_2\left|E^\prime,\kappa^\prime\right>=\kappa^*
\kappa^\prime\left<E,\kappa\right|z\left|E^\prime,\kappa^\prime\right>=-\left<E,\kappa\right|z\left|E^\prime,\kappa^\prime\right>
\end{eqnarray}
which leads to the expression
\begin{eqnarray}
  \left(\kappa^*\kappa^\prime+1\right)\left<E,\kappa\right|z\left|E^\prime,\kappa^\prime\right>=0.
\end{eqnarray}
Here we have used $\left<E,\kappa\right|\left(P_y P_z
I_{xy}S_2\right)^+=\left<\kappa\right|\kappa^*$. Apparently the
matrix element for $\pi$-transitions can only be non-zero for the
following combinations of $\kappa$ and $\kappa^\prime$:
\begin{eqnarray}
\pi:\quad(\kappa,\kappa^\prime)=(1,-1),(-1,1),(i,-i),(-i,i)
\end{eqnarray}
The above shows also that the expectation value of the $z-$coordinate
vanishes for any eigenstate i.e. we have $\left<E,\kappa|z|E,\kappa\right>=0$
For $\sigma^\pm$-transition one obtains in a similar way
\begin{eqnarray}
\sigma^+:\quad(\kappa,\kappa^\prime)&=&(i,1),(1,-i),(-1,i),(-i,-1)\\
\sigma^-:\quad(\kappa,\kappa^\prime)&=&(-i,1),(1,i),(-1,-i),(i,-1).
\end{eqnarray}
\begin{figure}[htb]\center
\includegraphics[angle=0,width=7cm]{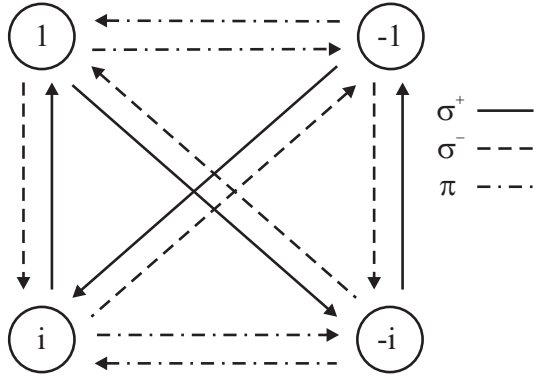}
\caption{Graphical representation of allowed dipole transitions
between $\kappa$-subspaces. The arrows point from $\kappa$
to $\kappa^\prime$.} \label{fig:selection_rules}
\end{figure}
Figure \ref{fig:selection_rules} presents an overview of
the allowed dipole transitions between the $\kappa$-subspaces.

We have calculated the dipole strengths for transitions from the
ground state to excited states.
\begin{figure}[htb]\center
\includegraphics[angle=0,width=7cm]{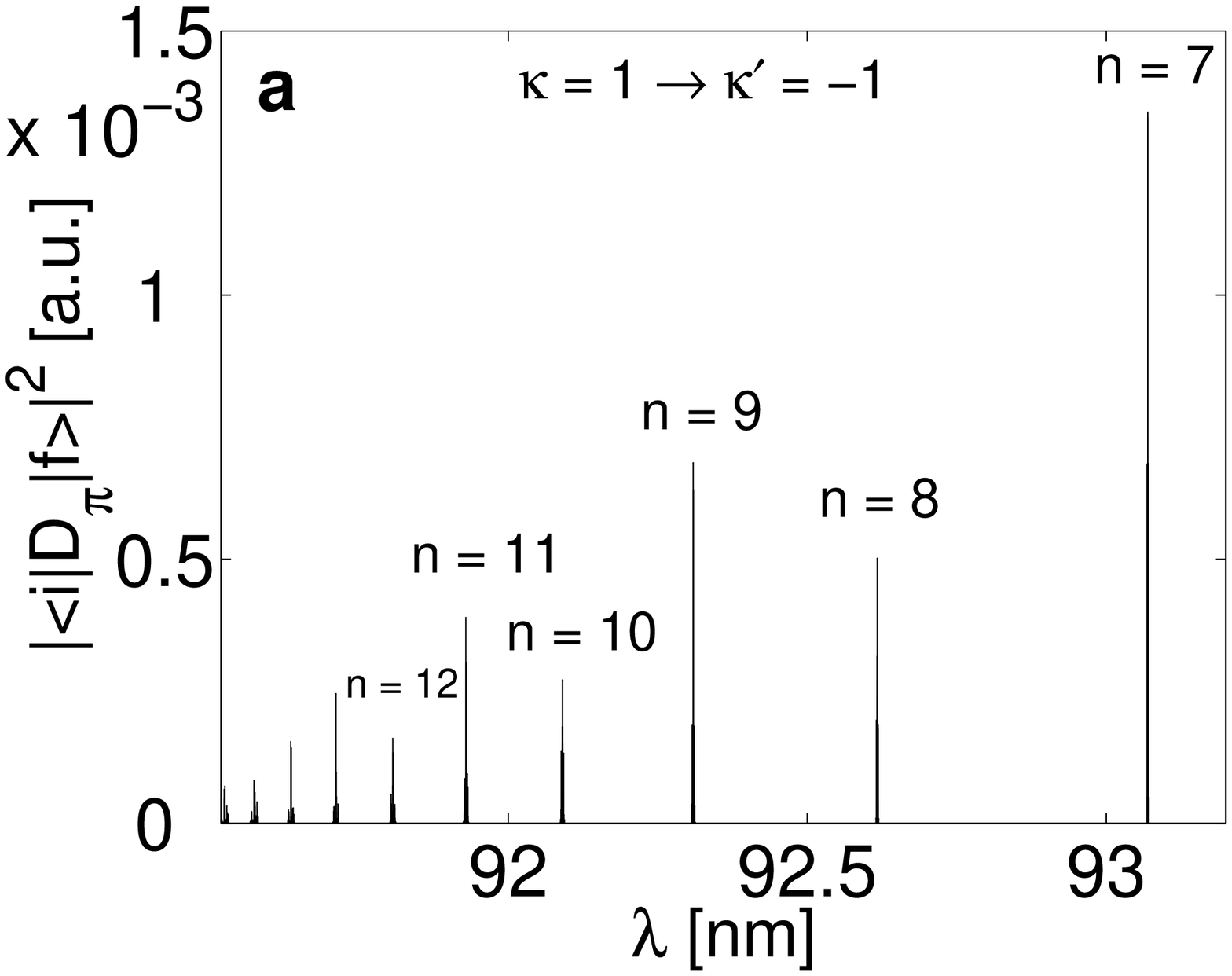}
\includegraphics[angle=0,width=7cm]{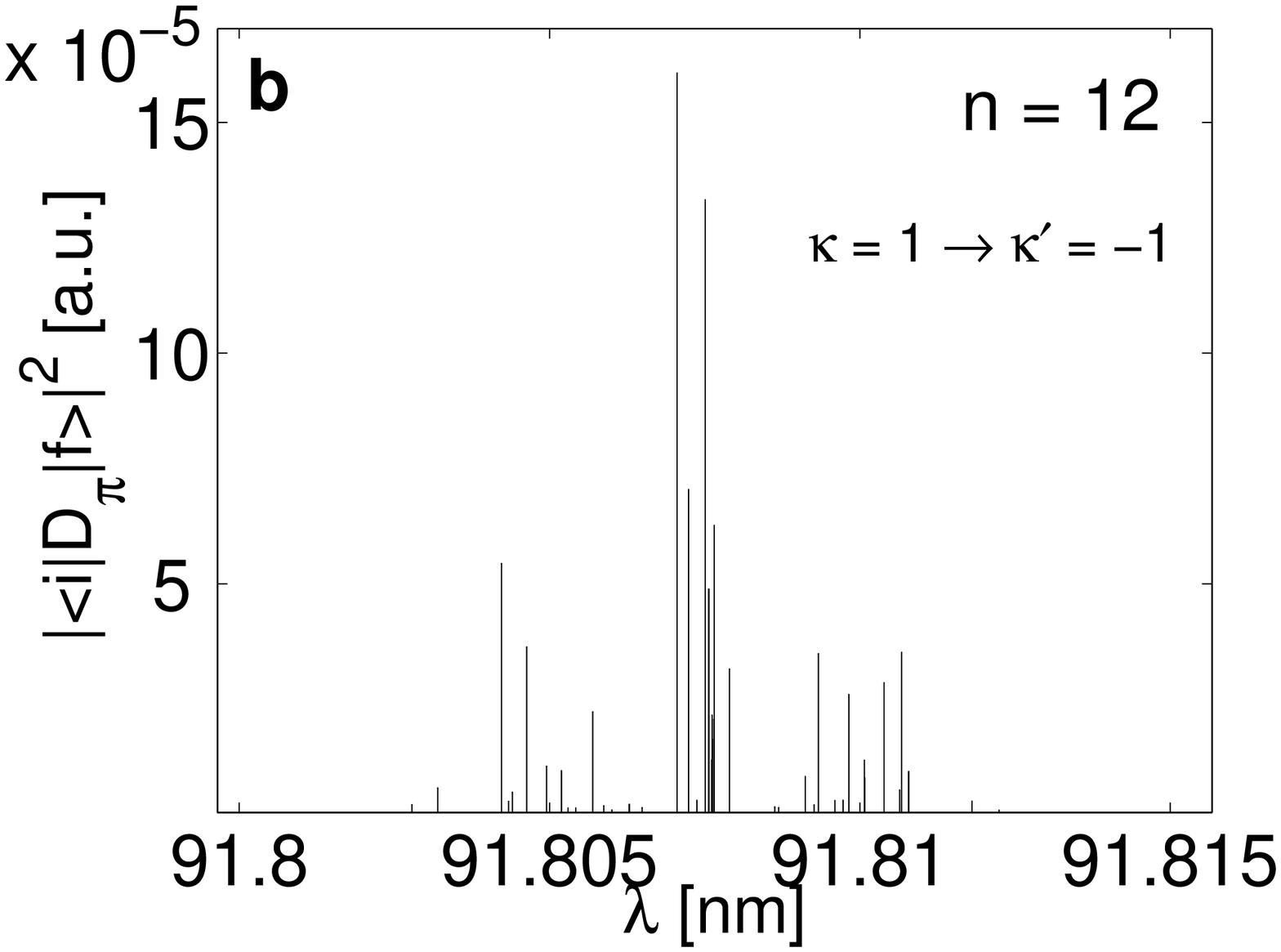}
\caption{\textbf{a:}  Dipole strenghts for $\pi$-transition from
the ground state of the $\kappa=1$-subspace to excited states
belonging to the $\kappa=-1$-subspace ($b=10^{-7}$). The line for
smallest $\lambda$ belongs to the $n=1\rightarrow 7$-transition.
\textbf{b:} Zoomed view of the line belonging to the transition to
the $n=12$-multiplet. The line center is dominated by two
sub-lines. The two bunches accompanying the line center at its
left and right hand side possess a much smaller dipole strength.}
\label{fig:pi_transitions}
\end{figure}
Figure \ref{fig:pi_transitions} shows the results we obtain for
$\pi$-transitions between the $\kappa=1$- and
$\kappa^\prime=-1$-subspace. In figure \ref{fig:pi_transitions}a
we observe a general decrease of the dipole strengths with
decreasing transition wavelengths. However, the decrease is not
monotonous as it would be in the case of a homogeneous or a
3D-quadrupole field \cite{Lesanovsky04_2}. One rather
finds a modulation on top of the transition amplitudes where the
$n=8$-, $n=10$- and $n=12$-multiplet exhibit smaller dipole
strengths than both of their neighbors. Figure \ref{fig:pi_transitions}b
shows a zoomed view of the $n=1 \rightarrow 12$ transition line.
Its structure is dominated by two sub-lines located in the line
center. The central bunch is almost symmetrically accompanied by
two bunches of sub-lines located for smaller and larger
wavelength, respectively.
\begin{figure}[htb]\center
\includegraphics[angle=0,width=7cm]{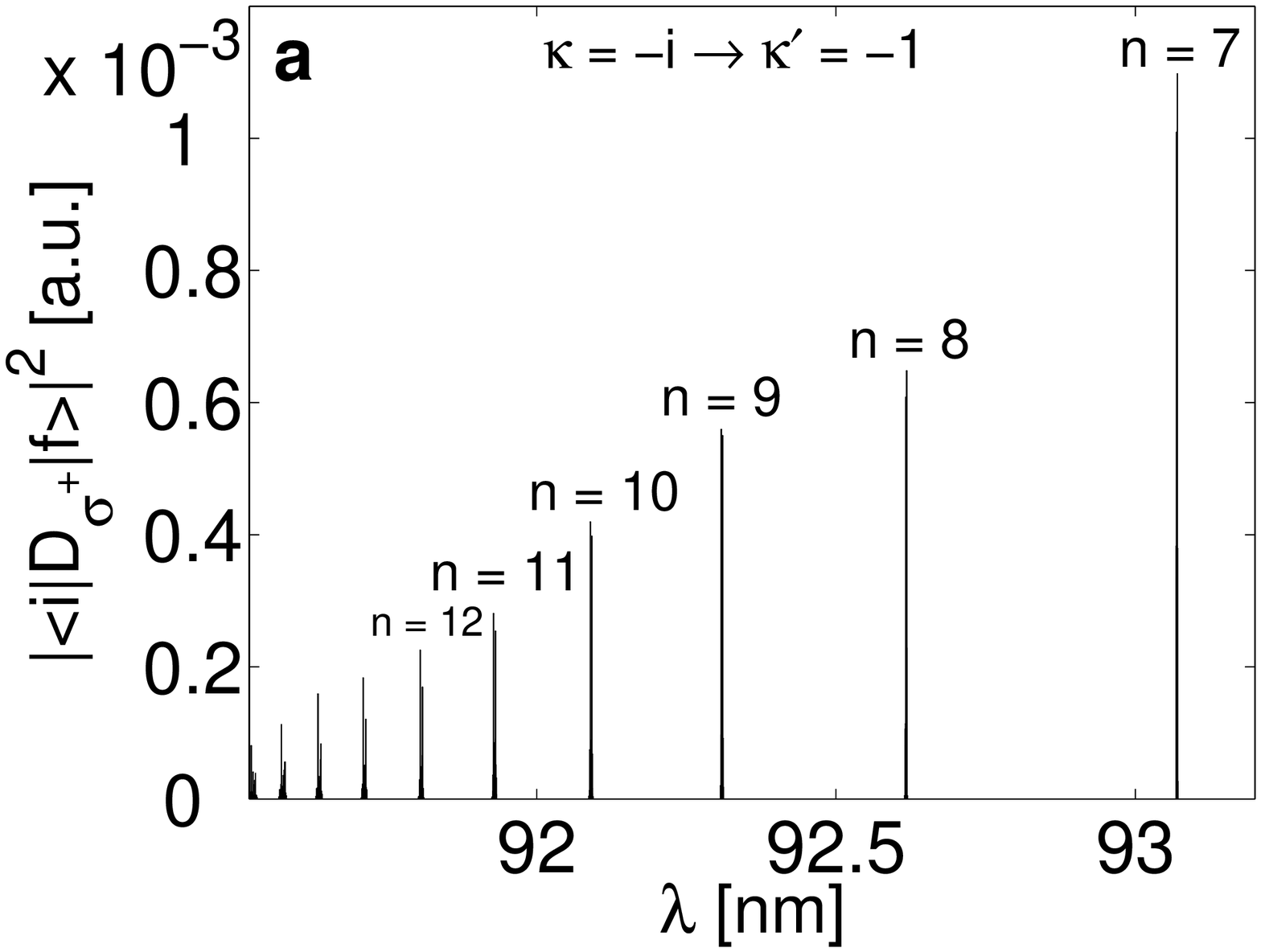}
\includegraphics[angle=0,width=7cm]{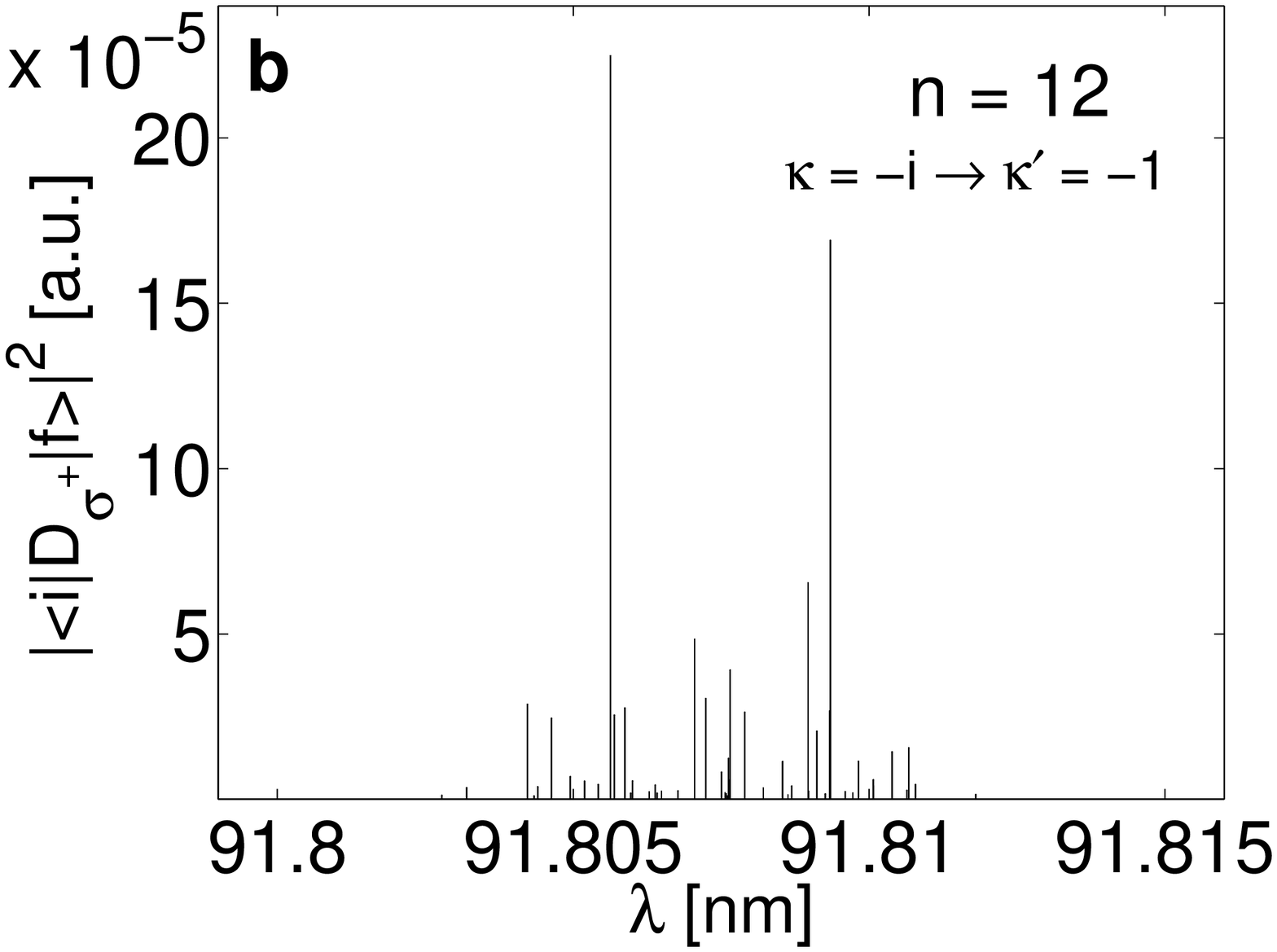}
\caption{\textbf{a:} Dipole strenghts for $\sigma^+$-transition
from the ground state of the $\kappa=-i$-subspace to excited
states belonging to the $\kappa=-1$-subspace ($b=10^{-7}$). The
line for smallest $\lambda$ belongs to the $n=1\rightarrow
7$-transition. \textbf{b:} Zoomed view of the line belonging to
the transition to the $n=12$-multiplet. The line consists of three
bunches each of which consisting of a number of sub-lines. The
line is dominated by two sub-lines one of each located in the left
and right hand side bunch.} \label{fig:sigma_transitions}
\end{figure}
For $\sigma^+$-transitions the dipole strengths are systematically
decreasing with decreasing wavelength (figure
\ref{fig:sigma_transitions}a). In the zoomed view (figure
\ref{fig:sigma_transitions}b) we also notice the structure
consisting of three bunches of sub-lines. Again there are two
dominating lines which are now located in the two outer bunches
rather than in the central one.

%
\subsection{Magnetic Guide with a Ioffe field} \label{subsec:ioffe}
%

As discussed in section \ref{sec:hamiltonian} an additionally
applied homogeneous field leads to severe changes of the symmetry
properties of the atomic system. Apart from the lifting of the
degeneracies also a significant influence on the electronic spin
and the transition amplitudes have to be expected.

Apparently there has to be a critical radius $\rho_c$ at which
both fields are equal in strength. For a given gradient $b$ and
homogeneous field strength $B_I$ it is given by
$\rho_c=\frac{B_I}{b}$. Taking into account the scaling
$\left<\rho\right>\propto n^2$ we expect states with
\begin{eqnarray}
  n_c=\sqrt{\frac{B_I}{b}}
\end{eqnarray}
to be equally affected by both fields. Hence, the states having
$n\ll n_c$ or $n\gg n_c$ should be dominated by the homogeneous field or
the field of the side guide, respectively.
\begin{figure}[htb]\center
\includegraphics[angle=0,width=7cm]{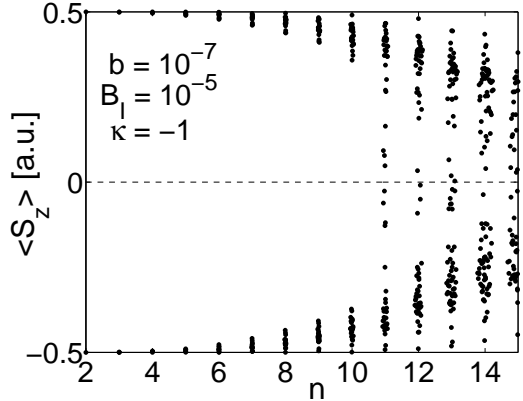}
\caption{Expectation values of the $z$-component of the electronic
spin at a finite homogeneous field strength ($B_I=10^{-5}$) and a
gradient of $b=10^{-7}$. At low degree of excitation the
homogeneous field dominates the electronic states. At this regime
$S_z$ becomes an approximate constant of motion admitting
$\left<S_z\right>$ to possess one of the two possible values
$\pm\frac{1}{2}$. States lying above the critical principal quantum
number $n_c$ become increasingly dominated by the quadrupole field.
As a result the expecation values tend towards
$\left<S_z\right>=0$.} \label{fig:sz_exp_value_IOFFE}
\end{figure}
Figure \ref{fig:sz_exp_value_IOFFE} shows the expectation values
of $S_z$ for a gradient $b=10^{-7}$ and a homogeneous field
strength $B_I=10^{-5}$. This yields the critical principal quantum
number $n_c=10$. Indeed one finds for $n\ll 10$ the expected
dominance of the homogeneous field. In this regime
$\left<S_z\right>$ is approximately allowed to possess one of the
two values $\pm\frac{1}{2}$. This is due to the fact that $S_z$
becomes an approximate constant of motion. For $n>10$ we observe
the expectation values to move towards zero which is expected from
the results shown in figure \ref{fig:sz_exp_value}. We have to
remark that since the symmetry $\Sigma_z$ persists the expectation
values of $S_x$ and $S_y$ vanish even for finite strength
of the homogeneous field.

Apart from the spin expectation value also the spin polarization
exhibits significant changes if a Ioffe field is switched on.
For a sufficient high field strength or low degree of
excitation $(n<n_c)$, respectively, the structure of the
electronic states is dominated by the Ioffe field. Here the spin
is expected to be aligned with the homogeneous field. Since $W_{SB}$
describes the projection of the electronic spin onto the direction
of the side guide field which is perpendicular to the Ioffe field
one expects $W_{SB}$ to be approximately zero in this regime.
\begin{figure}[htb]\center
\includegraphics[angle=0,width=5.2cm]{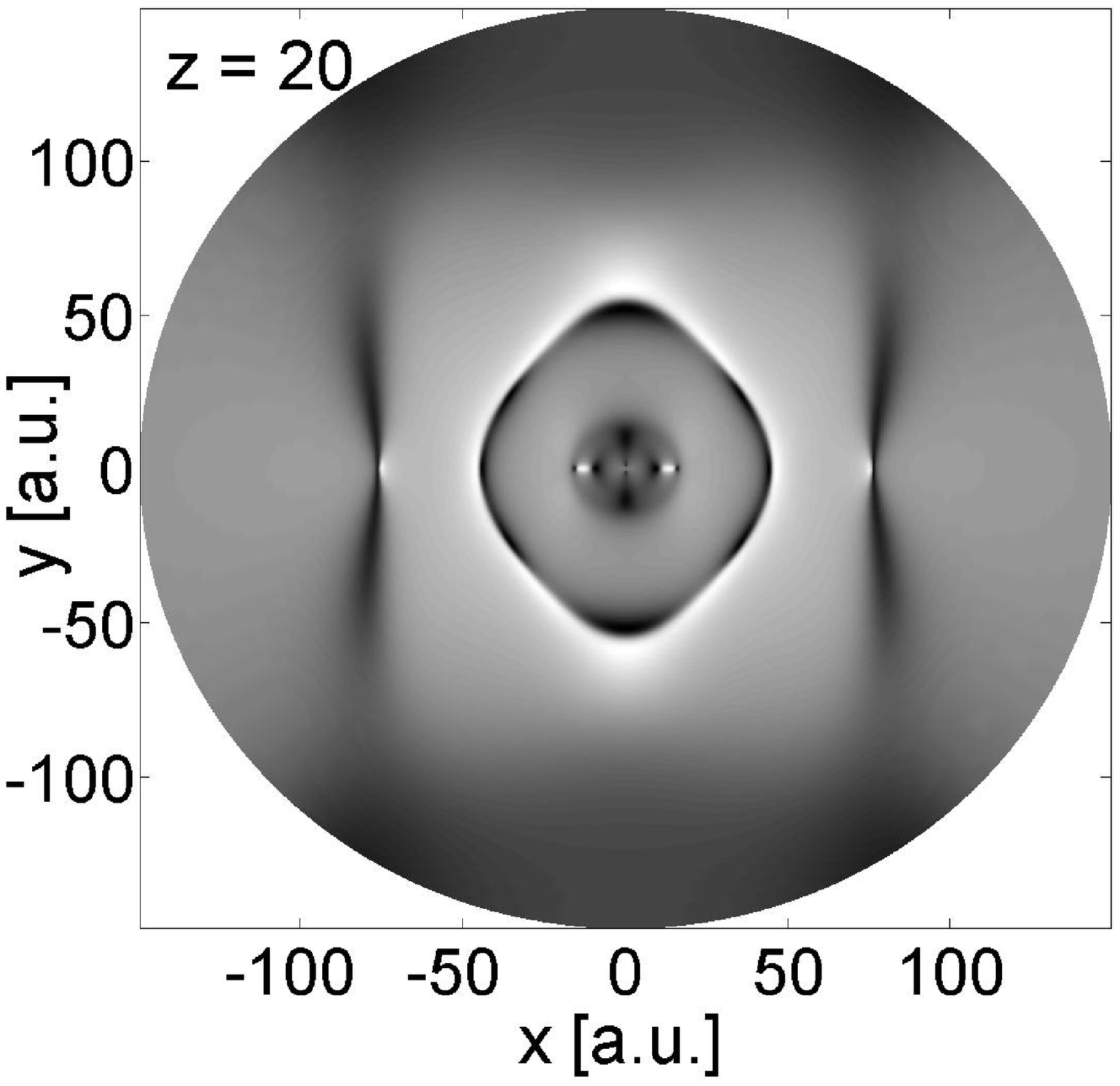}
\includegraphics[angle=0,width=5.2cm]{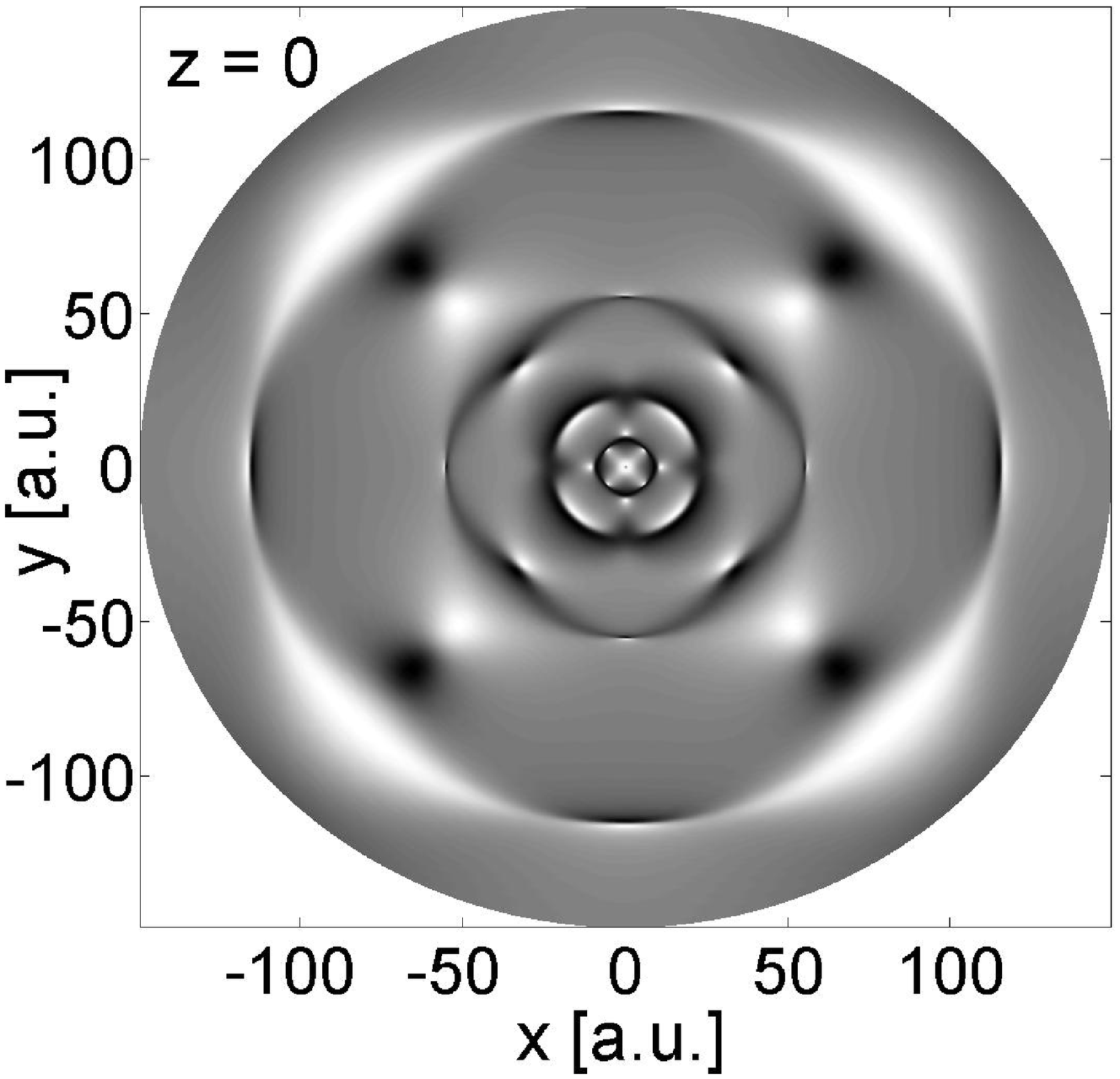}
\includegraphics[angle=0,width=5.2cm]{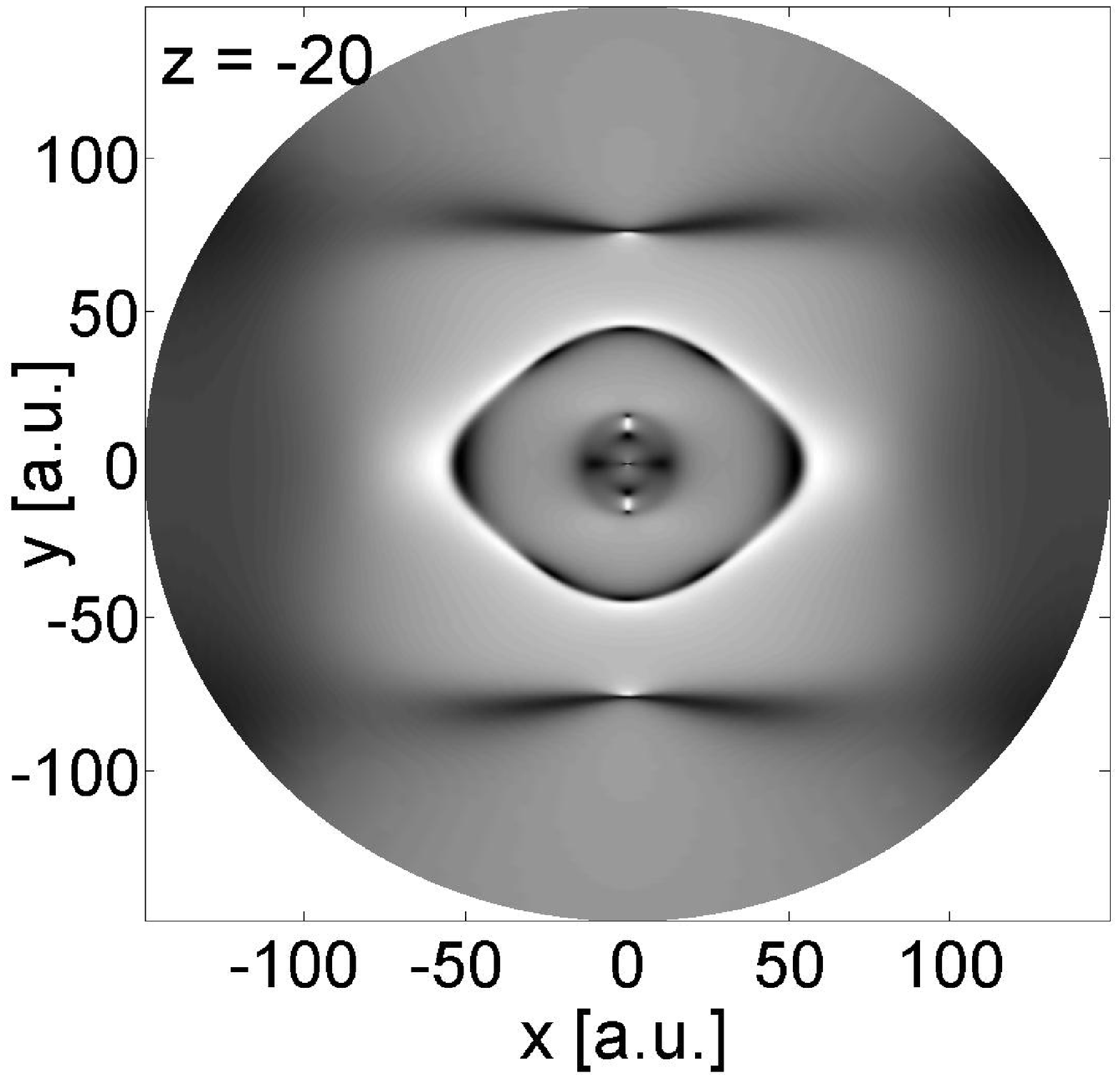}
\caption{Tomographic cuts through the spin polarization $W_{SB}$
(equation (\ref{eq:SB_polarization})) of the $83$rd excited state
at a finite Ioffe field strength $B_I=10^{-5}$. The state belongs
to the $n=8$ multiplet inside the $\kappa=1$-subspace
($b=10^{-7}$). The cuts are made at $z=\pm 20$ and $z=0$. Positive
and negative values are indicated by white and black, respectively.
One observes large gray regions with $W_{SB} \approx 0$.}
\label{fig:SB_density_ioffe}
\end{figure}
Figure \ref{fig:SB_density_ioffe} illustrates the $SB$
polarization $W_{SB}$ (equation (\ref{eq:SB_polarization})) for
the state shown in figure \ref{fig:SB_density} but for a Ioffe
field strength $B_I=10^{-5}$. The state is located inside the
$n=8$ multiplet which lies below the critical quantum number
$n_c=10$. Thus the states structure is dominated by the Ioffe
field. As expected from the discussion above we observe large gray
regions indicating $W_{SB}=0$. The geometry of the side guide field
is barely recognized for the cut made at $z=0$. Unlike in figure
\ref{fig:SB_density} there are only small regions exhibiting a
well-defined spin orientation that is dominated by the side guide, i.e either
$W_{SB}=-1$ or $W_{SB}=1$.

\begin{figure}[htb]\center
\includegraphics[angle=0,width=7cm]{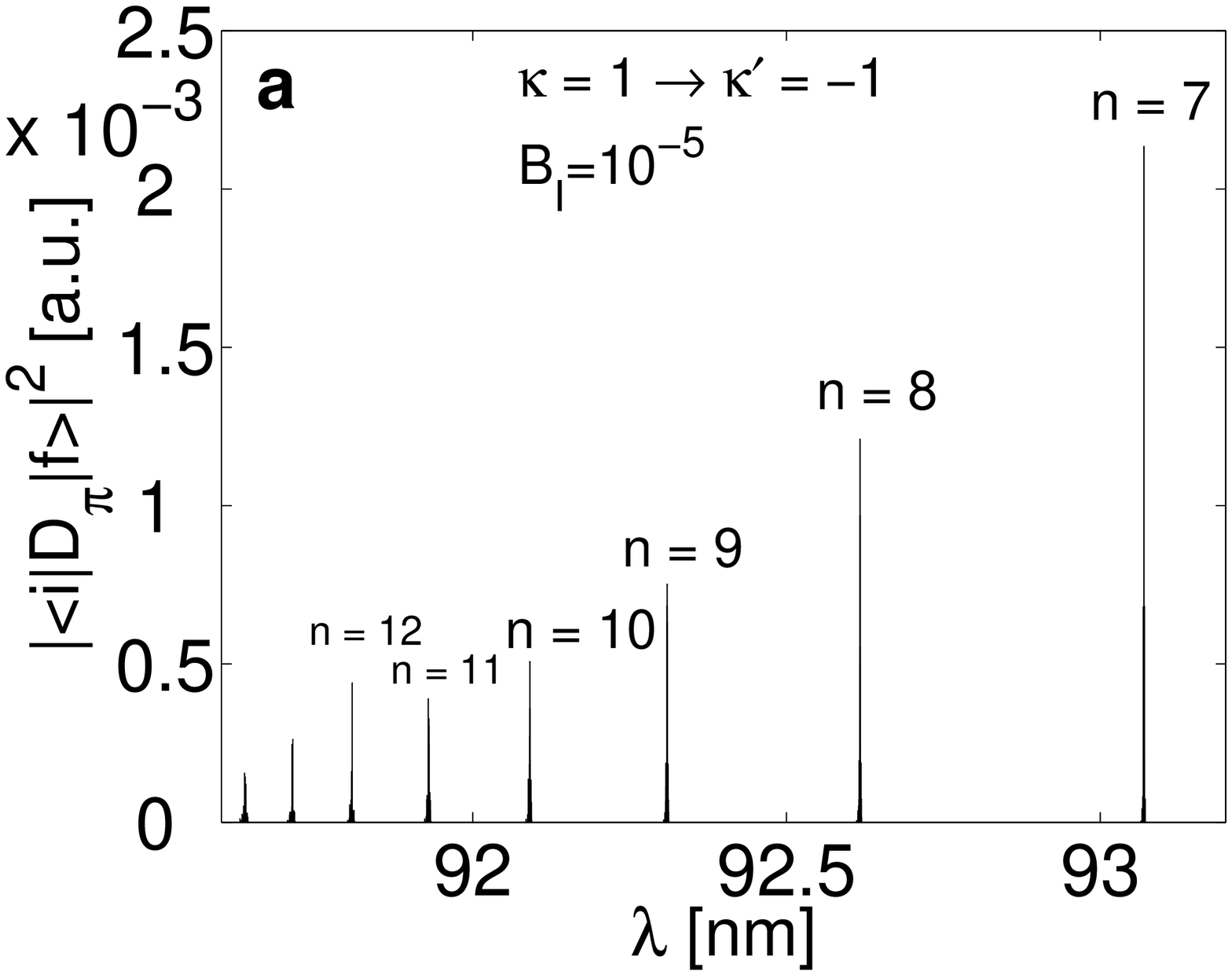}
\includegraphics[angle=0,width=7cm]{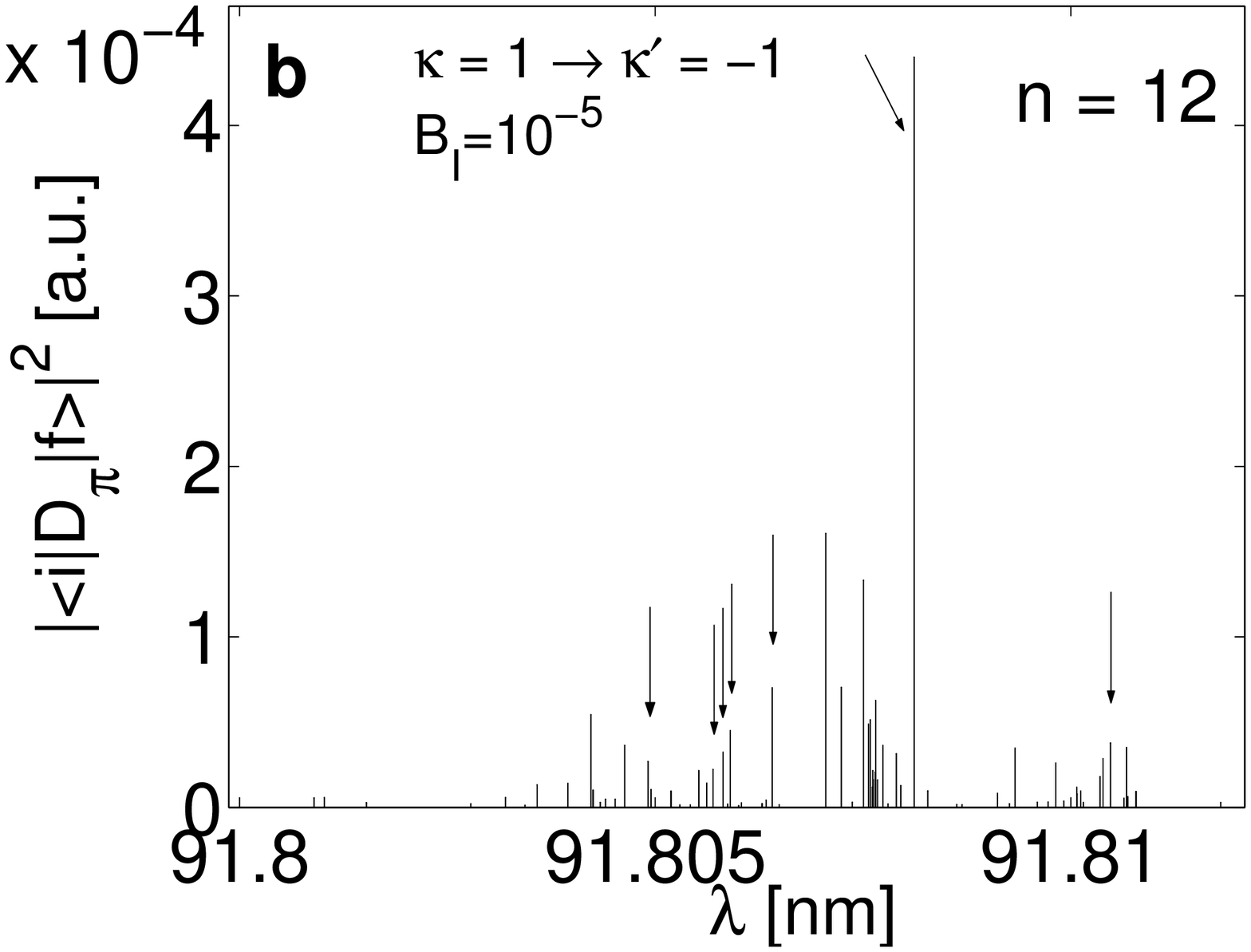}
\caption{\textbf{a:} Dipole strengths for $\pi$-transition from
the ground state of the $\kappa=1$-subspace to excited states
belonging to the $\kappa=-1$-subspace ($b=10^{-7}$ and
$B_I=10^{-5}$). The line for smallest $\lambda$ belongs to the $n=1\rightarrow
7$-transition. \textbf{b:} Zoomed view of the line belonging to
the transition to the $n=12$-multiplet. Several additional lines
appear at finite homogeneous field strength (some are marked by an
arrow). The line center is dominated by a single sub-line emerging
from a transition which is induced by the external homogeneous
field.} \label{fig:pi_transitions_IOFFE}
\end{figure}
Figure \ref{fig:pi_transitions_IOFFE}a shows the dipole strengths
for $\pi$-transitions from the ground state in the
$\kappa=1$-subspace to various states in the $\kappa=-1$-subspace.
Compared to the $B_I=0$ case the dipole strengths are increased by
approximately $70 \%$. The transition strengths increase with
increasing transition wavelengths. Again there seems to occur some
kind of modulation as already seen in figure
\ref{fig:pi_transitions}a but being less pronounced here. In the
present case the $n=12$-transition exhibits a larger transition
amplitude than its neighbors. In figure
\ref{fig:pi_transitions_IOFFE}b we show a zoomed view of the line
belonging to the $n=12$-transition. Due to the presence of the
homogeneous field a number of additional lines appear some of
which are marked by an arrow. In contrast to the $B_I=0$ case the
$n=12$ line is dominated by a single sub-line originating from a
transition induced by the presence of the homogeneous field.
%
\section{Conclusion and Outlook} \label{sec:outlook}
%
We have studied electronically excited hydrogen atoms located in a
magnetic guide. Including pseudo-potentials \cite{Gonzales03} for
the atomic core could be straight forwardly extended to describe
e.g. alkali atoms. The magnetic guide represents a microtrap used
to confine ultracold atomic systems. The motion of the valence
electron has been described by an effective one-body approach.
Both the coupling of the spatial degrees of freedom (para- and
diamagnetism) as well as the spin degrees of freedom to the
external field have been taken into account. The linear
variational principle has been used to solve the stationary
Schr\"odinger equation: Employing a Sturmian basis set enabled us
to converge a large number of eigenfunctions.

A careful inspection of the Hamiltonian yields an amazingly large
number of symmetries involving both the spin and spatial degrees
of freedom: We have found 15 symmetry operations of both unitary
and anti-unitary character. This allows for a classification of
the electronic eigenstates with respect to a complete set of
commuting constants of motion. The latter involve the Hermitian
$\Sigma_z$-operator which is a combined spin and parity operator
and the unitary but non-Hermitian operator $P_yP_zI_{xy}S_2$ which
involves parity and permutation operators. Employing specific
anticommuting operators of this symmetry group we could prove the
two-fold degeneracy of each energy level. This feature is indeed
shown to be generic for spin-$\frac{1}{2}$-systems exhibiting
certain symmetry properties. We have discussed how the symmetries
are affected if an additional homogeneous magnetic field is
applied in order to obtain a Ioffe-Pritchard type trap. In this
case only $7$ symmetry operations remain including $\Sigma_z$,
$P_yP_zI_{xy}S_2$ and $P_xP_zI_{xy}S_2^*$.

Spectra have been investigated up to energies corresponding to
a principal quantum number of $n\approx 15$. In the low gradient
regime degenerate $n$-manifolds split up symmetrically around the
zero field energy. For the intra-$n$-mixing regime only a very weak
restructuring takes place inside any $n$-multiplet i.e. we observe only
a minor nonlinear behaviour of the energies on the gradient. For even higher
gradients the inter-$n$-mixing takes place where states belonging to
adjacent multiplets begin to mix and avoided crossings dominate the spectrum.
Scaling relations for both, the inter- and the intra-$n$-mixing have been provided.

Effects due to the coupling of the spin and spatial degrees of freedom have been studied in detail.
An analysis of the spin-field orientation has been
performed by utilizing the distribution of the spin polarization. For
electronic states in the magnetic guide $W_{SB}$ reveals a rich
nodal and island structure which is absent for an atom in a uniform field.
Moreover an analysis of the $S_z$ expectation value has been performed. It has
been shown that states being energetically strongly affected by the presence of the magnetic guide
possess a small expectation value of $S_z$.

We have derived selection rules for the quantum number $\kappa$
belonging to the $P_yP_zI_{xy}S_2$ symmetry operator for linear as
well as circular polarized dipole transitions. Wave lengths and
dipole strengths from the ground to Rydberg states were analyzed.
In particular for $\pi$ transitions we have found a global
modulation of the transition amplitudes. The impact of the
presence of an additional homogeneous magnetic field (along the
wire involved in the set-up of the side guide) on several relevant
quantities has been studied. This includes the $S_z$-expectation
values and the electric dipole transition amplitudes.

Let us now comment on the approach chosen in the present work. Neglecting
the fine and hyperfine structure of the atom as well as omitting the influence of the core scattering events represent,
at least for certain species and regimes (high excitations !), certainly a good approximation to the true
physical system. Another approximation is the fact that we centered the nucleus at the minimum of the
field configuration. This is suggested by our assumption that we have ultracold atoms with an extremely
small kinetic c.m. energy in tight traps leading to a well-localized atomic c.m.
Nevertheless, it is expected that the c.m. motion blurs the effects ocurring for an atom with a fixed nucleus.
Beyond this, it is well-known that already in the presence of a homogeneous
magnetic field the c.m. and electronic motions of atoms do not separate i.e. they perform an intimately
coupled motion \cite{Avron78,Johnson83,Schmelcher94,Schmelcher92,Dippel94}. Then the immediate question
arises how this coupling might look like in our inhomogeneous field configuration and in particular what
its impact on the overall electronic motion is. To investigate this is a challenging task which needs careful
consideration and clearly goes beyond the scope of the present work.

\section{Acknowledgments}\label{sec:acknowledgments}

We are most thankful to Ofir Alon for fruitful discussions
regarding the group theoretical aspects of the present work. I.L.
acknowledges a scholarship by the
Landesgraduiertenf\"orderungsgesetz of the state of
Baden-W\"urttemberg.

\end{document}